\documentclass[twocolumn,amsmath,amssymb,superscriptaddress]{revtex4-1}

\usepackage[pdftex]{graphicx}
\usepackage{dcolumn}
\usepackage{bm}
\usepackage{amsmath, amssymb,amsfonts}
\usepackage{mathrsfs}
\usepackage{type1cm}
\usepackage[colorlinks,linkcolor=blue,citecolor=blue]{hyperref}

\begin{document}

\title{
Discrete-phase-space method for driven-dissipative dynamics of strongly interacting bosons in optical lattices 
}

\author{Kazuma Nagao}
\email{kazuma.nagao@riken.jp}
\affiliation{%
Computational Materials Science Research Team, RIKEN Center for Computational Science (R-CCS), Hyogo 650-0047, Japan
}%
\affiliation{%
Quantum Computational Science Research Team, RIKEN Center for Quantum Computing (RQC), Wako, Saitama 351-0198, Japan
}%
\author{Ippei Danshita}
\affiliation{%
Department of Physics, Kindai University, 3-4-1 Kowakae, Higashi-Osaka, Osaka 577-8502, Japan
}%
\author{Seiji Yunoki}
\affiliation{%
Computational Materials Science Research Team, RIKEN Center for Computational Science (R-CCS), Hyogo 650-0047, Japan
}%
\affiliation{%
Quantum Computational Science Research Team, RIKEN Center for Quantum Computing (RQC), Wako, Saitama 351-0198, Japan
}%
\affiliation{%
Computational Condensed Matter Physics Laboratory, RIKEN Cluster for Pioneering Research (CPR), Saitama 351-0198, Japan
}%
\affiliation{%
Computational Quantum Matter Research Team, RIKEN Center for Emergent Matter Science (CEMS), Saitama 351-0198, Japan
}%

\date{\today}%

\begin{abstract}

We develop a discrete truncated Wigner method to analyze the real-time evolution of dissipative SU(${\cal N}$) spin systems 
coupled with a Markovian environment. 
This semiclassical approach is not only numerically efficient but also particularly capable of accurately capturing local loss 
processes due to its local linearity in the dynamical equations.
We apply the method to a state-of-the-art experiment involving an analog quantum simulator of a three-dimensional dissipative 
Bose-Hubbard model in a strongly interacting regime. 
Our numerical results show good agreement with experimental data, specifically capturing the continuous quantum Zeno effect 
in the dynamics subjected to a gradual change of the ratio between the hopping amplitude and the onsite interaction across the 
superfluid-Mott insulator crossover. 
Furthermore, we present comparative analyses with the continuous truncated Wigner method, derived as an effective Fokker-Planck 
equation for SU(${\cal N}$) classical spin variables, showing that the discrete method outperforms the continuous one in simulating 
the long-time dynamics of SU(2) and SU(3) spin models.
The discrete phase space framework offers a versatile and powerful tool for exploring a wide range of open quantum many-body 
systems in dimensions higher than one dimension, where numerically exact methods are impractical due to the exponential growth 
of the Hilbert space dimension.

\end{abstract}

\maketitle

\section{Introduction}
\label{sec:1}

Rapid growth of technology for manipulating ultracold gases has enabled one to implement analog quantum simulators of open quantum many-body systems~\cite{diehl2008quantum,muller2012engineered,schafer2020tools,altman2021quantum}. 
Recent experiments have indeed realized interacting lattice systems with controllable dissipation, such as one-body losses~\cite{barontini2013controlling,labouvie2016bistability}, two-body losses~\cite{tomita2017observation,honda2023observation}, and inelastic photon scattering~\cite{patil2015measurement,luschen2017signatures,bouganne2020anomalous}.
These artificial systems have been utilized as platforms for exploring novel macroscopic states and dynamics of many-body systems beyond the paradigm of equilibrium and nonequilibrium physics of isolated systems. 
Recent experimental studies on Bose gases in optical lattices include the observations of bistability in steady states caused by one-body losses~\cite{labouvie2016bistability}, anomalous subdiffusion of superfluid coherence in the presence of both two-body losses and inelastic photon scattering~\cite{bouganne2020anomalous}, and the delay in melting of the Mott insulator due to the continuous quantum Zeno effects resulting from two-body losses~\cite{tomita2017observation}.
Such exploration of the nontrivial interplays between dissipation and many-particle correlations is relevant also to understanding and controlling various other systems, e.g., Bose-Einstein condensates coupled to optical cavity modes~\cite{brennecke2013real,chiacchio2019dissipation,kessler2021observation}, trapped ions~\cite{barreiro2011open,lin2013dissipative}, and superconducting qubits inside a cavity~\cite{leghtas2013stabilizing,chen2022decoherence}.

For further improvement of quantum simulators for open many-body systems, it is important to develop a numerical tool that can evaluate nontrivial outputs associated with the combined effects of interparticle interactions and dissipation.
However, reliable and efficient classical simulations of real time evolution of a dissipative many-body system is in general hard mainly due to the exponential growth of the Hilbert-space dimension with respect to the system size.
While some computational methods based on tensor networks have been developed and utilized~\cite{schroder2019tensor,luchnikov2019simulation,goto2020measurement,landa2020multistability,nakano2021tensor,mc2021stable,kilda2021stability}, their efficient applications are typically limited to low entangled states in one dimension as in the case of isolated systems.

For higher-dimensional open systems, there are various time-dependent mean-field approaches that can approximately evaluate the time dependence of local observables governed by microscopic master equations for mixed states.  
For instance, in Ref.~\cite{tomita2017observation}, the time-dependent Gutzwiller approximation method~\cite{diehl2010dynamical,tomadin2011nonequilibrium} is employed to analyze the time evolution of a dissipative Bose-Hubbard system with two-body losses.
It has been numerically shown that a site decoupling ansatz for many-body mixed states can nicely recover some qualitative features of the Zeno effect that can be observed in the local atomic density~\cite{tomita2017observation}.
However, in the Gutzwiller approach, for inducing crossover from a Mott insulator phase to a superfluid phase, one needs to introduce artificial noises as small perturbations.
Such noises inevitably cause uncontrolled errors, leading to quantitative inaccuracy in the results.
Moreover, due to its site decoupling nature, the Gutzwiller approach cannot properly describe offsite correlation functions. 
This drawback significantly limits the application of the Gutzwiller approach to ultracold-gas systems, e.g., not being able to evaluate the momentum distribution of atomic gases, an indispensable observable measured via time-of-flight images, because it is usually computed as a sum of offsite single-particle correlation functions in real space.

We also note that a dynamical mean-field theory (DMFT) approach based on a single-site impurity solver has been proposed to describe the quantum Zeno effect in higher dimensional dissipative Bose-Hubbard systems~\cite{scarlatella2021dynamical,secli2022steady}.
Even though the DMFT approach is expected to be more quantitative than the Gutzwiller approach, it is not straightforward to calculate spatial correlation functions either.
This difficulty stems from the locality of the formulation, in which the system sites are decomposed into an impurity site and its surrounding mean-field bath with only frequency dependence.

In this paper, we present truncated Wigner phase space methods for semiclassically analyzing dissipative dynamics of  strongly interacting Bose-Hubbard systems.
Phase space methods based on quasiprobability distributions such as the Wigner function, Husimi function, and Glauber-Sudarshan $P$ function provide a versatile tool to analyze open quantum systems~\cite{gardiner2004quantum,deuar2021multi,deuar2021fully}. 
In particular, the truncated Wigner approximation (TWA) in the Wigner function representation has recently emerged as a powerful semiclassical framework for approximately solving real time evolution of dissipative quantum systems coupled to the environment~\cite{vicentini2018critical,kessler2020continuous,seibold2020dissipative,hao2021observation,huber2021phase,vijay2022driven,mink2022hybrid,huber2022realistic}.
A generic feature of the TWA is that the dynamics of a quantum system is replaced with a set of fluctuating classical trajectories on phase space, whose initial conditions are statistically weighted with an initial state Wigner function expressing quantum fluctuations.

In a typical way of semiclassically expressing the dynamics of the isolated Bose-Hubbard model in its weakly interacting regime, trajectories are generated by the Hamilton equation for complex-valued field amplitudes, i.e., the discrete Gross-Pitaevskii (GP) equation~\cite{blakie2008dynamics,polkovnikov2010phase,nagao2019semiclassical,ozaki2020semiclassical}.
Although the TWA based on the GP equation no longer provides an appropriate description in the strongly interacting regime, there is an alternative representation, namely, the SU(${\cal N}$) TWA, which exactly describes onsite particle and hole fluctuations in strongly interacting states~\cite{davidson2015s,fujimoto2020family,nagao20213}.
In this method, the equation of motion in the classical limit is given as the Hamilton equation for SU(${\cal N}$) pseudospin variables, and it is formally equivalent to the time dependent variational equation in the single-site Gutzwiller mean-field theory~\cite{nagao20213}.
However, in the case of a driven dissipative system, which is not governed by the von-Neumann equation for isolated systems, the classical trajectories cannot be accurately reproduced via the Hamilton equations alone. 
This is because dissipative forces emerge in addition to the Hamiltonian dynamics, representing the effects of decoherence that influence the coherent time evolution.

As a key step to the goal of realizing an efficient large scale classical simulation of the experimental setup, here in this paper, we develop and use two types of the generalized SU(3) TWA method to analyze an effective Lindblad master equation for the dissipative analog quantum simulator. 
The first approach is a discrete TWA (dTWA) formulation~\cite{schachenmayer2015many,kunimi2021performance} for the master equation and gives a direct generalization of the SU(3) dTWA representation for the isolated Bose-Hubbard model, which was previously discussed in Ref.~\cite{nagao20213}.
The second one is based on a {\it continuous} TWA representation of the Markovian time evolution built with a stochastic Langevin equation and a continuous Wigner distribution for the initial state.
Since the Langevin equation derived here for the SU(3) TWA is inevitably nonlinear 
with respect to the dissipation strength in the master equation, 
such a description is merely approximate for any dissipation strength even if the offsite term, i.e., the hopping term in the Hamiltonian, 
is absent. 
By contrast, the classical equation of motion appearing in the dTWA formalism is linear with respect to the onsite dissipation strength, 
thus implying that the onsite two-body losses can be exactly captured independently of the dissipation strength. 
Therefore, an improved accuracy of the semiclassical description is anticipated in the first approach.
We note that the strategy of linearization can be readily implemented for arbitrary SU(${\cal N}$) spin systems, as demonstrated for a dissipative single SU(2) spin in Appendix~\ref{LZ model}, and would be powerful for a situation where a many-body system is subjected to local dissipation.

We demonstrate that these two different TWA approaches can qualitatively describe the continuous quantum Zeno effect 
in the nonunitary dynamics with strong dissipation.
However, the resulting threshold of the dissipation strength for exhibiting the Zeno effect is different, 
depending on which approach is employed for the simulation. 
In particular, we show that, because of the linearity of the nonunitary contributions in the classical limit, the dTWA representation can produce reasonable threshold values of the dissipation strength separating Zeno and non-Zeno regimes in comparison with the experimental results on the time evolution of the atomic density as well as the momentum distribution.
The local linearity of the dTWA allows it not only to predict reasonable values of the Zeno threshold but also to provide valuable 
insights into the qualitative behavior of long-time loss dynamics.
Indeed, the dTWA approach can capture second-order perturbation processes of hopping in the original Bose-Hubbard representation, 
as manifested by an emergent scaling law for atomic density decay after rescaling with an appropriate time unit. 
Additionally, we show that the dTWA approach can qualitatively reproduce the power-law decay in the long-time evolution of 
atomic density, with the extracted exponents exhibiting a dependence on the dissipation strength similar to the exact results.
These features cannot be described by the FP-TWA approach.

The rest of this paper is organized as follows.
In Sec.~\ref{sec:2}, we introduce an effective Lindblad master equation for onsite two-body losses of atoms in the quantum simulator.
In Sec.~\ref{sec:3}, we describe two different SU(3) TWA formalisms for the master equation, which are not equivalent to each other 
in the presence of dissipation.
In Sec.~\ref{sec:4}, we show the numerical results to investigate the continuous Zeno effect. 
In Sec.~\ref{sec:conclusions}, we conclude this paper and describe future perspectives. 
In Appendix~\ref{LZ model}, the dTWA methods are applied to a dissipative SU(2) spin system.  
A complete set of SU(3) base matrices is described in Appendix~\ref{def:matrices}. 
The derivation of the dTWA for an SU(3) spin system is detailed in Appendix~\ref{app:dtwa}, and  
a numerical sampling scheme for phase-space trajectories is explained in Appendix~\ref{app:sample}. 
Finally, in Appendix~\ref{app:fptwa}, details of the Fokker-Planck TWA are provided.

\section{Model}
\label{sec:2}

The system considered in this paper is an open quantum many-body system with controllable dissipation of local inelastic two-body losses.
The relevant experimental system is a cold atomic analog quantum simulator reported in Ref.~\cite{tomita2017observation}, where 
a highly-controlled dissipative Bose-Hubbard system is successfully realized, as outlined below.

\begin{figure}
\begin{center}
\includegraphics[width=80mm]{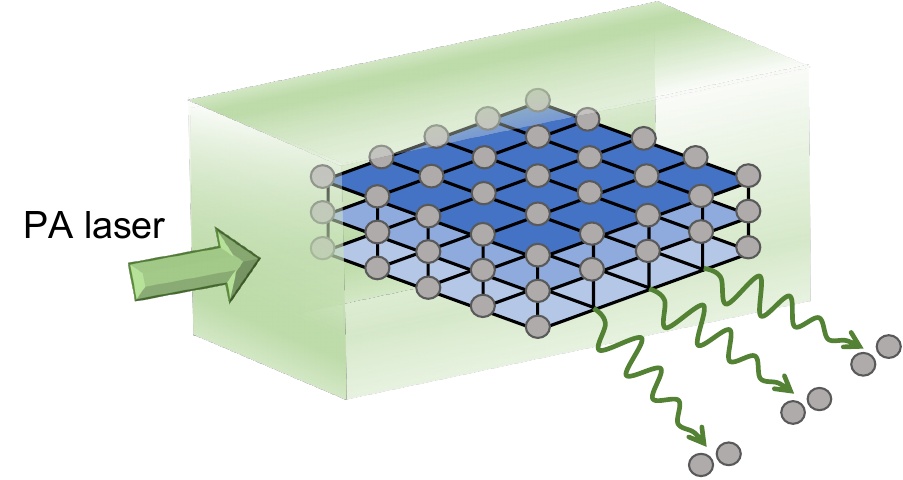}
\vspace{-2mm}
\caption{
Schematic illustration of the analog quantum simulation experiment for a driven-dissipative Bose-Hubbard system in Ref.~\cite{tomita2017observation}.
In order to effectively induce two-body loss dissipation, a PA laser is shined to an atomic system comprised of $^{174}{\rm Yb}$ atoms in a three-dimensional optical lattice.
}
\label{fig:PAlaser}
\end{center}
\end{figure}

In Ref.~\cite{tomita2017observation}, Tomita {\it et al.} utilized a single-photon photoassociation (PA) laser technique to realize a controllable dissipative system built with $^{174}{\rm Yb}$ bosonic atoms in a three dimensional optical lattice.
Shining a PA laser beam to this optical lattice system may convert pairs of two atoms doubly occupying the same site into ${}^{1}S_{0} + {}^{3}P_{1}$ molecules~\cite{tomita2017observation}. 
These molecules are immediately disassociated into the two ground-state atoms with high kinetic energy, 
and thus these disassociated atoms escape from the system, as schematically illustrated in Fig.~\ref{fig:PAlaser}.
An important point of this setup is that the strength of such onsite inelastic two-body collision loss can be readily controlled 
via tuning the intensity of PA laser.

The atomic system that is coherently coupled to the photo-associated molecular state is described by the Markovian-Lindblad master equation with a one-body loss Kraus operator for the molecules~\cite{tomita2017observation,rousseau2008quantum}.
However, our focus is on the real-time dynamics of the atomic sector of this atom-molecule coupled system.
We therefore define an effective density matrix operator ${\hat \rho}_{\rm eff}$ only for the atoms by adiabatically erasing the molecular degrees of freedom from the total density matrix of the composite system~\cite{tomita2017observation}.
This step can be carried out by using second order perturbation theory with some projection operators for the master equation.
The time evolution equation for ${\hat \rho}_{\rm eff}$ is then found to be an effective Lindblad master equation that only includes the 
atomic field operators, i.e., 
\begin{align}
\hbar \frac{\partial}{\partial t}{\hat \rho}_{\rm eff} = - i [ {\hat H}_{\rm BH}, {\hat \rho}_{\rm eff} ] + L_2[{\hat \rho}_{\rm eff}]. \label{eq: lindblad}
\end{align}
The first commutator term expresses the unitary contribution to the dynamics generated by the Bose-Hubbard Hamiltonian in three dimensions~\cite{jaksch1998cold}
\begin{align}
{\hat H}_{\rm BH} = -J\sum_{\langle j,k \rangle}({\hat a}^{\dagger}_{j}{\hat a}_{k}+{\rm H.c.})+\frac{U}{2}\sum_{j}{\hat a}^{\dagger}_{j}{\hat a}^{\dagger}_{j}{\hat a}_{j}{\hat a}_{j}, \label{eq:bhh}
\end{align}
where ${\hat a}_{j}$ and ${\hat a}^{\dagger}_{j}$ are the bosonic annihilation and creation operators at site $j$, respectively, 
satisfying the commutation relations $[{\hat a}_{j},{\hat a}^{\dagger}_{l}]=\delta_{j,l}$.   
$J$ and $U$ represent the hopping amplitude and the onsite interaction strength, respectively, and they are defined as functions of the optical lattice depth that can be readily controlled in the experimental setup.

In Eq.~(\ref{eq: lindblad}), the dissipative contribution in the presence of the PA laser is described by the nonunitary term $L_2[{\hat \rho}_{\rm eff}]$.
For the setup that we consider here, the onsite losses of atoms are represented by 
\begin{align}
L_2[{\hat \rho}_{\rm eff}]= \frac{\hbar \Gamma_{\rm PA}}{4}\sum_{j}\left[2{\hat a}_{j}{\hat a}_{j}{\hat \rho}_{\rm eff}{\hat a}^{\dagger}_{j}{\hat a}^{\dagger}_{j}-\{{\hat a}^{\dagger}_{j}{\hat a}^{\dagger}_{j}{\hat a}_{j}{\hat a}_{j},{\hat \rho}_{\rm eff}\} \right], 
\nonumber
\end{align}
where $\{\cdot,\cdot\}$ is the anti-commutator bracket.
The inelastic collision rate $\Gamma_{\rm PA}$ is expressed as $\Gamma_{\rm PA} = \beta_{\rm PA} \int d{\bf r} |w({\bf r})|^4$, where $w({\bf r})$ is the lowest band Wannier function as a function of the lattice depth and $\bf r$ is the coordinate vector in real space.
The coefficient $\beta_{\rm PA}$ can be controlled by varying the intensity of the PA laser, as described above  
(for more details, also see Ref.~\cite{tomita2017observation}).
We note that the same term describing two-body losses emerges in an optical lattice system of metastable $^{174}$Yb atoms in the $^{3}P_{2}$ state as well~\cite{tomita2019dissipative}.
However, the strength of dissipation, corresponding to $\Gamma_{\rm PA}$ described above, is not controllable in this case because it is directly determined by the intrinsic collision rate in the metastable state.

The controllability of the collision rate $\Gamma_{\rm PA}$ has been utilized in Ref.~\cite{tomita2017observation} to explore the continuous quantum Zeno effect on coherent many-body dynamics in the strongly interacting regime of the three-dimensional lattice system.
This effect occurs as suppression of coherent tunneling processes between neighboring sites caused by strong dissipation of losses~\cite{zhu2014suppressing,patil2015measurement,tomita2017observation,rossini2021strong,asai2022transition,honda2023observation,rosso2022one}.
The quantum Zeno effect in many-body systems has also been an important concept in recent studies associated with 
measurement-induced entanglement transitions on quantum circuits~\cite{li2019measurement,koh2023measurement}.
In the following, we analyze driven-dissipative dynamics of the experimental system by using SU(3) TWA methods for the master equation in Eq.~(\ref{eq: lindblad}).
As a key consequence demonstrating the effectiveness of the SU(3) TWA, we will show that this approximation can 
qualitatively capture the dissipation-induced delay of the coherent time evolution, i.e., the continuous Zeno effect, observed experimentally 
in the atomic density and the momentum distribution of the system~\cite{tomita2017observation}.

\section{Methods}
\label{sec:3}

\subsection{
General remarks
}

Before proceeding, we briefly make key remarks on the SU(${\cal N}$) TWA method that we employ.
This method is a phase-space approach for numerically simulating real-time dynamics of interacting spin systems 
for spin $S \geq 1/2$~\cite{wurtz2018cluster}.
Within the approximation, the time evolution of a quantum spin model reduces to an ensemble of fluctuating trajectories 
on the phase space of classical SU(${\cal N}$) spins.
The stochastic nature arises from the presence of quantum fluctuations in the initial state.
The idea of the SU(${\cal N}$) TWA relies on the fact that any local non-linear operator in spin-$S$ systems can be mapped 
to a linear combination of generators of the SU(${\cal N}$) group, where ${\cal N} = 2S+1$.
For instance, for an $S=1$ magnet with single-ion uniaxial anisotropy, its quadratic interaction term, i.e., $\sum_{i}({\hat S}^{z}_{i})^2$, 
is transformed into a linear term of generators of the SU(3) group. 
The TWA applied to such a linearized version of the model significantly improves beyond the capability of the naive approximation 
for the original model written in terms of the SU(2) spin operators, i.e., ${\hat S}^{x}_{i}$, ${\hat S}^{y}_{i}$, and ${\hat S}^{z}_{i}$~\cite{davidson2015s}.
In previous studies, the SU(${\cal N}$) TWA has been applied to experiments for magnetic atoms to realize a large-$S$ spin system~\cite{lepoutre2019out}, spin-1 systems with single-ion uniaxial anisotropy and strongly interacting Bose-Hubbard systems~\cite{davidson2015s,fujimoto2020family,nagao20213}, and cluster representations of spin-$1/2$ systems in one dimension~\cite{wurtz2018cluster}.

Even though the following analyses are devoted to the ${\cal N}=3$ case, our strategy can be readily generalized to any open 
quantum system in the presence of local dissipation terms in arbitrary spatial dimensions.
In Appendix~\ref{LZ model}, we demonstrate this with a single-site spin-1/2 model coupled with a Markovian bath.

\subsection{SU(3) matrix representation of the dissipative Bose-Hubbard system}

To apply the SU(${\cal N}$) TWA to the effective Lindblad master equation in Eq.~(\ref{eq: lindblad}), we derive a reduced model in a projected Hilbert space spanned by relevant Fock states. 
In the experiment, the onsite interaction $U$ is typically tuned to be sufficiently large compared to the hopping amplitude $J$, i.e., $U \gg J$.
This condition implies that the relevant single-particle states of onsite occupation are only a few states around a mean filling state~\cite{altman2002oscillating,nagao2018response,nagao20213}.
In this paper, we assume that the mean filling state at each site is the unit filling state $|1\rangle$, and the relevant onsite excitations around it are the vacuum state $|0\rangle$ and the doubly occupied state $|2\rangle$, therefore with ${\cal N}=3$.
In such a reduced space, written as $\bigotimes_j\{|0\rangle,|1\rangle,|2\rangle \}_{j}$, any physical operator, including the Hamiltonian operator and the effective density matrix, can be represented as an expansion in tensor products of three-dimensional SU(3) base matrices such as the Gell-Mann matrices~\cite{georgi2000lie}.

Let us define SU(3) pseudospin operators ${\hat X}^{(j)}_{\alpha} = ({\hat X}^{(j)}_{\alpha})^{\dagger}$ $(\alpha=1,\cdots,8)$ acting 
on the reduced Hilbert space.
Each of these operators has a finite-dimensional tensor product representation as
\begin{align}
{\hat X}^{(j)}_{\alpha} \Leftrightarrow I \otimes \cdots I \otimes \underbrace{T_{\alpha}}_{j\text{\rm th}} \otimes I \otimes \cdots, 
\end{align}
where eight Hermitian matrices $T_{\alpha}$ form a complete set of SU(3) base matrices, 
explicitly defined in Appendix~\ref{def:matrices}, 
and $I$ is the three-dimensional unit matrix.
The base matrices are assumed to satisfy the $su(3)$ Lie algebras $[T_{\alpha},T_{\beta}]=if_{\alpha\beta\gamma}T_{\gamma}$ with a structure factor tensor $f_{\alpha\beta\gamma}$, and the repeated Greek indices imply Einstein's summation convention. 
With this definition, the SU(3) operators recast the commutation relations $[{\hat X}^{(j)}_{\alpha},{\hat X}^{(l)}_{\beta}]=if_{\alpha\beta\gamma}{\hat X}^{(j)}_{\gamma}\delta_{j,l}$.

In the reduced Hilbert space, each annihilation operator is represented as a linear combination of the SU(3) operators, i.e.,
\begin{align}
{\hat a}_{j} 
&\rightarrow |0 \rangle\langle 1 |_{j} + \sqrt{2} |1 \rangle\langle 2 |_{j} \nonumber \\
&\;\;\; = p_{1}{\hat X}_{1}^{(j)} + i p_{1} {\hat X}_{2}^{(j)} + p_{2} {\hat X}_{6}^{(j)} + i p_{2} {\hat X}_{7}^{(j)} \nonumber \\
&\;\;\; \equiv {\tilde a}_{j}. \label{eq: a_eff}
\end{align}
The linear coefficients of Eq.~(\ref{eq: a_eff}) are then uniquely determined as $p_{1}=\frac{1+\sqrt{2}}{2\sqrt{2}}$ 
and $p_{2}=\frac{\sqrt{2}-1}{2\sqrt{2}}$.
Moreover, because of the linearization property of $T_{\alpha}$, implying that any polynomial of $T_{\alpha}$ can be recasted into a linear combination of $T_{\alpha}$ and $I$, the quadratic particle density operator ${\hat n}_{j} = {\hat a}^{\dagger}_{j}{\hat a}_{j}$ becomes 
\begin{align}
{\hat n}_{j} \rightarrow 1 - {\hat X}_3^{(j)},
\end{align}
and the onsite interaction operator ${\hat a}^{\dagger}_{j}{\hat a}^{\dagger}_{j}{\hat a}_{j}{\hat a}_{j}$ is represented as 
\begin{align}
{\hat a}^{\dagger}_{j}{\hat a}^{\dagger}_{j}{\hat a}_{j}{\hat a}_{j} \rightarrow \frac{2}{3} - {\hat X}^{(j)}_{3} - \frac{1}{\sqrt{3}}{\hat X}^{(j)}_{8}. 
\end{align}

Combining these results, ${\hat H}_{\rm BH}$ is replaced with a SU(3) spin Hamiltonian~\cite{nagao20213}
as 
\begin{align}
{\hat H}_{\rm BH}
&\rightarrow {\hat H}_{\rm BH}^{\rm SU(3)} =  - \frac{U}{2\sqrt{3}}\sum_{i}{\hat X}^{(i)}_{8} - \frac{U}{2}\sum_{i}{\hat X}^{(i)}_{3} \nonumber \\
&\;\;\;\;\;\;\;\;\;\;\;   -J \sum_{\langle i,j \rangle}\left( {\tilde a}^{\dagger}_{i} {\tilde a}_{j} + {\rm H.c.}\right). \label{eq:effective_hamiltonian}
\end{align}
${\hat H}_{\rm BH}^{\rm SU(3)}$ is composed of a linear term of the SU(3) pseudospin operators, which is proportional to $U$, and a nonlinear term, which is proportional to $J$.
If $J = 0$, the unitary time evolution governed by the von-Neumann equation for Eq.~(\ref{eq:effective_hamiltonian}) is exactly reproduced in the TWA associated with the linearity in $U$~\cite{nagao20213}.
For $J\neq 0$, but $U \gg J$, the corresponding Hamilton equation with the Weyl symbol $H_{\rm BH}^{W} = ({\hat H}_{\rm BH}^{\rm SU(3)})_W$, which is explicitly provided in Appendix~\ref{app:dtwa}, is a nonlinear equation of motion with weak nonlinearity.
Hence, the semiclassical description of the quantum dynamics is expected to be reasonable over a valid timescale, which can be longer than that for the conventional TWA with the GP equation~\cite{nagao20213}.

For applications to the effective Lindblad master equation in Eq.~(\ref{eq: lindblad}), the dissipative $L_2$ term also has 
to be represented in the pseudospin operators.
Because of the linearization property of the base matrices, the two-body loss operator ${\hat a}^2_{j}$ for a doubly occupied site, 
which provides the Kraus measurement operator in constructing $L_2 [{\hat \rho}_{\rm eff}]$, also reads as a linear combination of 
${\hat X}_{4}^{(j)}$ and ${\hat X}_{5}^{(j)}$, i.e., 
\begin{align}
{\hat a}^2_{j}  \rightarrow \frac{1}{\sqrt{2}}{\hat X}_{4}^{(j)} + i \frac{1}{\sqrt{2}}{\hat X}_{5}^{(j)}. 
\end{align}
Note that offsite dissipative terms spreading over multiple sites, which arise, e.g., in a dissipative hard-core boson system in Ref.~\cite{garcia2009dissipation}, cannot be straightforwardly linearized in onsite phase space variables of spins.

\subsection{Method I: dTWA}

Let us now explain the dTWA~\cite{schachenmayer2015many,orioli2017nonequilibrium,kunimi2021performance,vijay2022driven} to approximately calculate the time evolution of the dissipative interacting SU(3) spin system.
The dTWA method for finite levels or spin systems formally uses a phase point operator to expand the density matrix operator.
For the case of SU(3) spin systems with $N$ sites, an appropriate definition of phase point operator is given via its matrix representation, i.e., ${\mathscr A}^{\otimes N}=\bigotimes_{j=1}^N{\mathscr A}^{[j]}$ 
with ${\mathscr A}^{[j]}=\frac{1}{3}I +\frac{1}{2}{\bm r}^{(j)} \cdot {\bm T} $~\cite{nagao20213}~and the normalization conditions of $T_{\alpha}$ being ${\rm Tr}(T_{\alpha}T_{\beta})=2\delta_{\alpha \beta}$.
Moreover, ${\bm r}^{(j)} = (r^{(j)}_{1},\cdots,r^{(j)}_{8})$ represents a spinor vector of eight real-valued discrete parameters.
Using the phase point operator, one can expand the initial density matrix at $t=0$ such that $\rho (t=0) = \prod_{j}\sum_{{\bm r}^{(j)}} {\mathscr W}({\bm r}^{(1)},\cdots,{\bm r}^{(N)}) {\mathscr A}^{\otimes N}[{\bm r}^{(1)},\cdots,{\bm r}^{(N)}]$.
The expansion coefficient ${\mathscr W}({\bm r}^{(1)},\cdots,{\bm r}^{(N)}) $ is interpreted as a discrete Wigner function associated with the phase point operator.
For a direct product state over real space, the discrete Wigner function is given as a spatially factorized distribution function, i.e., ${\mathscr W}({\bm r}^{(1)},\cdots,{\bm r}^{(N)}) = \prod_{j}{\mathscr W}_{1}^{[j]}({\bm r}^{(j)})$.
The local distribution ${\mathscr W}_{1}^{[j]}({\bm r}^{(j)})$ depends only on ${\bm r}^{(j)}$. 
The explicit form of ${\mathscr W}_{1}^{[j]}({\bm r}^{(j)})$ used in the main calculations is provided in Appendix~\ref{app:sample}.

With the initial density matrix expanded with the phase point operator described above, 
we now consider the nonunitary time evolution of the phase point operator, which is also generated by the effective Lindblad master equation in Eq.~(\ref{eq: lindblad}).
Let us introduce a time-dependent phase point operator at time $t$ as ${\hat {\mathscr A}}^{\otimes N}_t [\{{\bm r}^{(k)}\}_{k=1}^{N}]$ 
with the initial condition ${\hat {\mathscr A}}^{\otimes N}_{t=0} = {\hat {\mathscr A}}^{\otimes N}$.
This operator has to evolve according to 
\begin{align}
\hbar \frac{\partial}{\partial t} {\hat {\mathscr A}}^{\otimes N}_t = -i [{\hat H}_{\rm BH}^{\rm SU(3)},{\hat {\mathscr A}}^{\otimes N}_t] + L_2[{\hat {\mathscr A}}^{\otimes N}_t]. \label{eq: eom_ppo}
\end{align}
In the dTWA, to obtain a mean-field equation that generates classical trajectories, a direct product ansatz is assumed 
as a solution of this equation~\cite{schachenmayer2015many}, which is expressed as 
${\mathscr A}^{\otimes N}_t [\{{\bm r}^{(k)}\}_{k=1}^{N}] = \bigotimes_{j} {\mathscr A}_{j,{\rm dTWA}}^{(1)}(t)$ with 
${\mathscr A}_{j,{\rm dTWA}}^{(1)} (t)=  \frac{1}{3}I +\frac{1}{2}{\bm r}^{(j)}_{\rm cl}(t;\{{\bm r}^{(k)}\}_{k=1}^{N})\cdot {\bm T}$. 
The time dependent parameters ${\bm r}^{(j)}_{\rm cl}(t)$, which are generally nonlinear functions of their initial values ${\bm r}^{(j)} = {\bm r}^{(j)}_{\rm cl}(t=0)$, behave as a set of phase space variables propagated with a classical equation of motion.
Quantum fluctuations in the initial state are incorporated in the discrete Wigner function ${\mathscr W}$.
In numerical simulations, this Wigner function makes a statistical ensemble of ${\bm r}^{(j)}_{\rm cl}(t)$ to produce dynamical formation of offsite spatial correlations~\cite{kunimi2021performance}.
Technical details to derive the dTWA are supplemented in Appendix.~\ref{app:dtwa}.

Hereinafter, we identify the time evolution equation of ${\bm r}^{(j)}_{\rm cl}$.
The variables ${\bm r}^{(j)}_{\rm cl}$ are nothing but the discrete Weyl symbols of the SU(3) spin operators defined via the phase point operator, i.e., $r^{(j)}_{{\rm cl},\alpha} = {\rm Tr} \left[ T_{\alpha} {\mathscr A}^{(1)}_{j,{\rm dTWA}}(t) \right]$.
From this, we obtain a nonlinear differential equation in the general form
\begin{align}
{\dot r}_{{\rm cl},\alpha}^{(j)} = - i\hbar^{-1}\{r^{(j)}_{{\rm cl},\alpha},H_{\rm BH}^{W}\}_{\rm P.B.} + F_{\alpha}^{(j)}[{\bm r}^{(j)}_{\rm cl}]. \label{eq: dissipative_eom_dtwa}
\end{align}
The Poisson bracket for the SU(3) spins can be defined as the symplectic operator acting on classical numbers 
$\{ \cdot , \cdot \}_{\rm P.B.} = i \sum_{j} \frac{\overleftarrow \partial}{\partial r^{(j)}_{{\rm cl},\alpha}}f_{\alpha\beta \gamma} r^{(j)}_{{\rm cl},\gamma} \frac{\overrightarrow \partial}{\partial r^{(j)}_{{\rm cl},\beta}}${\color{blue}~\cite{davidson2015s}}.
Note that the commutator contribution $\{r^{(j)}_{{\rm cl},\alpha},H_{\rm BH}^{W}\}_{\rm P.B.}$ in Eq.~(\ref{eq: dissipative_eom_dtwa}) 
gives rise to nonlinear interactions between spins at neighboring sites. 
For isolated systems, the differential equation in Eq.~(\ref{eq: dissipative_eom_dtwa}) reduces to the Hamilton equation to generate trajectories under conservative forces.
However, in the presence of the onsite two-body losses, there is an additional contribution to the equation 
described by dissipative forces $F_{\alpha}^{(j)}$.
These forces turn out to be linear functions of the phase-space variables, which are specified as 
\begin{align}
F_{1}^{(j)}
&= F_{6}^{(j)} = - \frac{\Gamma_{\rm PA}}{4}(r_{{\rm cl},1}^{(j)}+r_{{\rm cl},6}^{(j)}), \nonumber \\
F_{2}^{(j)}
&= F_{7}^{(j)} = - \frac{\Gamma_{\rm PA}}{4}(r_{{\rm cl},2}^{(j)}+r_{{\rm cl},7}^{(j)}), \nonumber \\
F_{3}^{(j)}
&= \Gamma_{\rm PA}\left(\frac{2}{3} - r_{{\rm cl},3}^{(j)} - \frac{1}{\sqrt{3}}r_{{\rm cl},8}^{(j)} \right), \label{eq: linear_force} \\
F_{4}^{(j)}
&= - \frac{\Gamma_{\rm PA}}{2} r_{{\rm cl},4}^{(j)}, \;\;\;
F_{5}^{(j)} = - \frac{\Gamma_{\rm PA}}{2} r_{{\rm cl},5}^{(j)}, \nonumber \\
F_{8}^{(j)}
&= 0. \nonumber
\end{align}
The linearity of the $\Gamma_{\rm PA}$ term in Eq.~(\ref{eq: linear_force}) implies that the local loss processes, e.g., 
those where two atoms occupying the same site escape from the system, can be exactly captured in the dTWA.
In Sec.~\ref{exact}, we demonstrate numerically the exact solvability of the dTWA in the case of the site decoupling limit at $J=0$.

Within the dTWA, the quantum expectation value of an observable represented by an operator ${\hat {\cal O}}$ at time $t$, 
$\langle {\hat {\cal O}}(t) \rangle$, 
is given by a phase-space average, i.e., 
\begin{align}
\langle {\hat {\cal O}}(t) \rangle = \left\langle {\cal O} [{\bm r}^{\alpha}_{\rm cl}(t;{\bm r}^{\beta})] \right\rangle^{\mathscr W}_{{\bm r}^{\beta}}, 
\end{align}
where the bracket symbol $\langle \cdots \rangle^{\mathscr W}_{{\bm r}^{\beta}}$ in the righthand side is an ensemble average over possible 
configurations of ${\bm r}^{\beta}$ 
and ${\cal O}[{\bm r}^{\alpha}_{\rm cl}(t;{\bm r}^{\beta})]$ is a classical function corresponding to ${\hat {\cal O}}$.
The time dependence of the phase-space average is due to the classical trajectories ${\bm r}^{\alpha}_{\rm cl}(t;{\bm r}^{\beta})$, which are the solutions of Eq.~(\ref{eq: dissipative_eom_dtwa}) for snapshots of ${\bm r}^{\beta}$ distributed according to the discrete Wigner function ${\mathscr W}$.
In this study, we numerically solve Eq.~(\ref{eq: dissipative_eom_dtwa}) by using a simple integration scheme of nonlinear ordinal differential equations, i.e., the explicit fourth-order Runge-Kutta scheme.
In addition, for the Monte Carlo evaluation of the average, we utilize a quantum-state-tomography inspired method, namely the generalized discrete Wigner (GDW) sampling scheme~\cite{zhu2019generalized,nagao20213} described in Appendix~\ref{app:sample}.

As characterized by the onsite linearity, the SU(3) dTWA approach has advantages over other phase space methods, 
such as the positive-$P$ approach~\cite{deuar2021fully}, in keeping track of long-time evolution of strongly interacting open systems.
Indeed, in the dTWA, no instability occurs in the sampling of trajectories for arbitrary parameters, provided that the time grid is chosen 
to be sufficiently small. 
However, in the positive-$P$ approach, a kind of {\it noise amplification} in trajectories occurs around a finite time threshold, 
leading to uncontrolled instabilities in quantum expectation values. 
Although such instabilities can be alleviated by adding a dissipation term to the dynamics, as demonstrated for a single-body loss 
model in Ref.~\cite{deuar2021fully}, the long-time evolution in the strongly interacting regime, which is relevant to the melting dynamics 
of a Mott insulator, might be difficult to access in this approach.
In addition, while trajectories in the SU(3) dTWA can capture transitions between superfluid and Mott-insulating states in a 
site-decoupling approximation~\cite{nagao20213}, trajectories in the positive-$P$ representation cannot, as its classical equation 
falls into a class of GP-type equations~\cite{gardiner2004quantum}.
In open systems with weak interactions, where the valid timescale of the SU(3) dTWA is quite shortened, the positive-$P$ approach 
can be stable over relatively long times and might have potential advantages in quantitative simulations.

\subsection{Method II: Fokker-Planck TWA}
\label{subsec: FPTWA}

As described above, the local linearity is essential for the dTWA to be able to capture qualitative behavior of the dissipative dynamics 
caused by onsite two-body losses.
To better understand the effectiveness of such a linearity in the classical limit, for comparison, here we also introduce more 
standard {\it continuous} TWA formulation with the form of the Fokker-Planck (FP) equation for the continuous Wigner function of 
the effective density matrix operator $\rho_{\rm eff}$~\cite{gardiner2004quantum,huber2021phase}.
We call this approach the FP-TWA, hereafter.
For interacting spin systems, the FP-TWA is formulated on the basis of the continuous phase-space representation in terms of Schwinger's 
constrained bosons with multi flavors or orthogonal sets of Hermitian bilinear operators in such bosons, i.e., the Jordan-Schwinger mapping of the 
special unitary group generators~\cite{polkovnikov2010phase,wurtz2018cluster,huber2021phase}.

To derive a FP equation from Eq.~(\ref{eq: lindblad}), we generalize a Bopp operator approach for the bilinear operators~\cite{polkovnikov2010phase,wurtz2018cluster}.
For the operator set of ${\hat X}_{\alpha}^{(j)}$, Bopp operators truncated to leading order are given by ${\hat X}_{\alpha}^{(j)} \rightarrow x_{\alpha}^{(j)} + \frac{i}{2}f_{\alpha\beta\gamma}x_{\gamma}^{(j)}\frac{\overrightarrow \partial}{\partial x_{\beta}^{(j)}}$. These allow one to obtain 
an effective FP equation for the distribution $W_{\rm eff}=W_{\rm eff}(\{x_1^{(j)},\cdots,x_8^{(j)}\})$ with a positive semidefinite diffusion matrix $D = {\rm diag}_{j}[(D^{(j)}_{\alpha\beta})]$~\cite{gardiner2009stochastic,gardiner2004quantum},
\begin{align}
\frac{\partial W_{\rm eff}}{\partial t} 
&= \sum_{j}{\cal L}^{(j)}_{\rm FP}(W_{\rm eff}),  \label{eq: fokker-planck} 
\end{align}
where 
\begin{align}
{\cal L}^{(j)}_{\rm FP}( \bullet ) 
&= -\frac{\partial}{\partial x_{\alpha}^{(j)}} (f_{\alpha}^{(j)} \bullet ) + \frac{1}{2}\frac{\partial^2}{\partial x_{\alpha}^{(j)} \partial x_{\beta}^{(j)}}(D^{(j)}_{\alpha \beta} \bullet) \nonumber 
\end{align}
is a local FP operator to propagate the Wigner function and  
the first (second) term of the right hand side is called the drift (diffusion) term.

The FP equation can be solved by integrating a stochastic Langevin equation with the general form 
\begin{align}
d x_{\alpha}^{(j)} = f_{\alpha}^{(j)} dt + \sum_{q=1}^{2}g_{\alpha,q}^{(j)} \cdot dw_{q}^{(j)}, \label{eq: langevin}
\end{align}
where $\alpha = 1,2,\cdots,8$ and $q = 1,2$.
The drift term is translated into the deterministic forces $f_{\alpha}^{(j)} dt = (f_{{\rm unitary},\alpha}^{(j)}+f_{{\rm PA},\alpha}^{(j)})dt$, including nonunitary contributions $f_{{\rm PA},\alpha}^{(j)}$ from the PA laser, while the diffusion term gives rise to the stochastic forces $\sum_{q=1}^{2}g_{\alpha,q}^{(j)} \cdot dw_{q}^{(j)}$.
In the limit of $\Gamma_{\rm PA} = 0$, Eq.~(\ref{eq: langevin}) reduces to the Hamilton equation for SU(3) variables~\cite{nagao20213}, i.e., $d x_{\alpha}^{(j)} = -(i/\hbar) \{  x_{\alpha}^{(j)}, H_{\rm BH}^{W} \}_{\rm P.B.} dt = f_{{\rm unitary},\alpha}^{(j)} dt $.
The local noises $dw_{q}^{(j)} = {\cal O}(dt^{1/2})$ 
must obey a Gaussian probability distribution around the zero mean value, whose variance is specified by the normalization conditions $dw_{q}^{(j)} dw_{q'}^{(j')} = dt \delta_{q,q'}\delta_{j,j'}$.
In addition, the dot product between $g_{\alpha,q}^{(j)}$ and $dw_{q}^{(j)}$ has to be interpreted as the It{\^o} product for random variables~\cite{gardiner2009stochastic}.
The analytical expressions of $f_{{\rm PA},\alpha}^{(j)}$ and $g_{\alpha,q}^{(j)}$ are provided in 
Eqs.~(\ref{eq:f_PA}) and (\ref{eq:g}), respectively, in Appendix~\ref{app:fptwa}, which can recover the original FP equation.
In particular, the results of $g_{\alpha,q}^{(j)}$ are consistent with the factorization property of the diffusion matrix, i.e., $D^{(j)}=B_jB^{\rm T}_j$, where $B_j = [{\bm g}_{q=1}^{(j)},{\bm g}_{q=2}^{(j)},{\bm 0},\cdots,{\bm 0}]$ is an $8 \times 8$ matrix.

The force vectors arising in the presence of the PA laser are given as nonlinear functions of $x_{\alpha}^{(j)}$, 
and the nonlinearity is measured by $\Gamma_{\rm PA}$.
The FP-TWA for the Lindblad master equation is therefore not exact even in the zero hopping limit $J=0$, 
implying that the valid timescale of the approximation becomes shorter as $\Gamma_{\rm PA}$ increases.
We will numerically compare the results obtained by the FP-TWA with those obtained by the dTWA 
to examine the limitations of the FP-TWA in detail in Sec.~\ref{exact}.
When $\Gamma_{\rm PA} = 0$ and $J = 0$, the Langevin equation reduces to a linear equation, 
and the FP-TWA becomes exact, as expected. 
Therefore, for dissipative systems, these two approaches, i.e., the Bopp operator method and the phase point operator method, 
lead to noticeable differences in the presence of nonunitary contributions.

In the FP description, the quantum expectation value of an operator ${\hat {\cal O}}$ at time $t$ reads as 
\begin{align}
\langle {\hat {\cal O}}(t) \rangle = \left\langle \left\langle {\cal O}_{W} \left[ \{ x^{(j)}_{{\rm cl},\alpha}(t) \} \right] \right\rangle \right\rangle_{\rm Langevin + Initial \; noise}. \label{eq: TWA_FP}
\end{align}
The doubled brackets indicate that the average includes Gaussian noises during the classical time evolution as well as the initial 
noises drawn from the Wigner function at $t=0$.
The details of numerically evaluating the average are explained in Appendix~\ref{app:fptwa}.

\section{Numerical simulations}
\label{sec:4}

We now apply the generalized SU(3) TWA methods described in the preceding sections to the dissipative Bose-Hubbard system 
in three dimensions. 
We numerically demonstrate that both methods can describe the continuous quantum Zeno effect in the nonunitary 
many-body dynamics qualitatively within the accuracy of the semiclassical representations.
To reduce the total runtime of the TWA simulations, we parallelize the sampling of random phase-space 
variables at $t=0$.
The parallelization, done with the function of message passing interface (MPI), allows a number of processes, 
as many as $n_{\rm MPI} = 20\;000$ processes, to run simultaneously in a single computational job. 
The task of calculating a single trajectory from an initial stochastic realization is assigned to each MPI process. 
If a large number of Monte Carlo samples, e.g., $\chi$ samples with $\chi \gg 10^{4}$, are required for convergence, each MPI process 
is then assigned to generate $n_{\chi}$ trajectories and compute a partial average of an observable over them, 
where $n_{\chi} > 1$ and $n_{\chi} \times n_{\rm MPI} = \chi$.

\subsection{Stability of the unit filling Mott insulating state}
\label{subsec:quench}

\begin{figure*}[!t]
\begin{center}
\includegraphics[width=180mm]{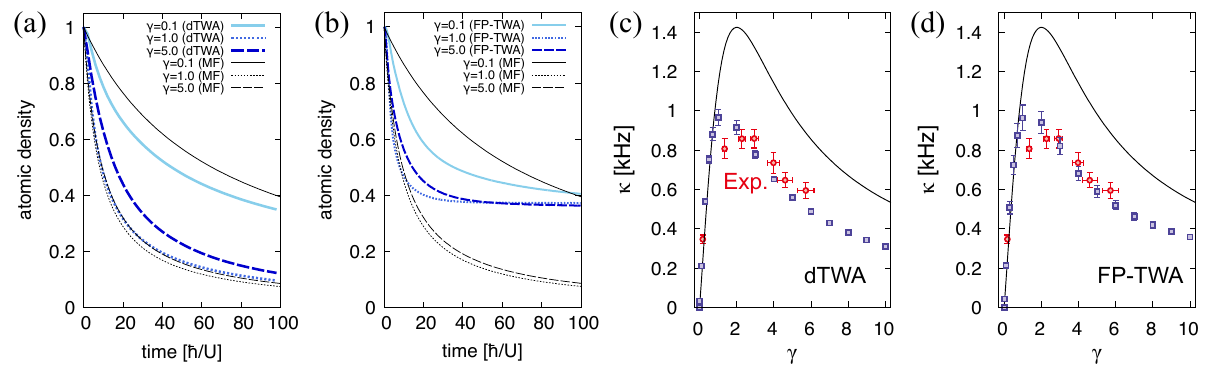}
\vspace{-6mm}
\caption{
(a) 
SU(3) dTWA results for the nonunitary time evolution of the atomic density $n_j(t)$ starting from the unit filling Mott-insulating state 
$|\psi_{\rm Mott}\rangle$ at time $t=0$ for three different values of the dissipation strength $\gamma = \hbar \Gamma_{\rm PA}/U$.
The results for $\gamma = 0.1$, $1$, and $5$ are shown by bold solid, dotted, and dashed lines, respectively.
The corresponding mean-field results with the form $(1+\kappa_{\rm MF}t)^{-1}$ are also plotted as thin black solid, dotted, and dashed lines, respectively. 
(b)
Same as (a) but for the SU(3) FP-TWA results.
(c) 
Two-body loss rate $\kappa$ as a function of $\gamma$ (blue open squares) extracted from the dTWA results 
for $n_j(t)$ shown in (a).
The error bar for each result denotes the standard estimation error for the fitting.
We select the data points of density $n_j(t)$ in (a) for the fitting such that the distance between adjacent data points on the time axis 
is $\delta t_{\rm fit} = 2.0 \hbar/U$.
The black solid line is the mean-field result $\kappa_{\rm MF}$ of a dissipative hard-core boson model.
(d) 
Same as (c) but for the SU(3) FP-TWA results. 
For comparison, the experimental results reported in Ref.~\cite{tomita2017observation} are also plotted as red open circles 
in (c) and (d). 
Here, in all calculations, we set the hopping amplitude $J=0.08U$.
}
\label{fig:loss_rate}
\end{center}
\end{figure*}

As explored experimentally in Ref.~\cite{tomita2017observation}, we study the stability of the unit filling Mott insulating state 
in the presence of the two-body loss dissipation by using the semiclassical methods described above.
We numerically study the time evolution of the dissipative Bose-Hubbard system given in Eq.~(\ref{eq: lindblad}) 
at a fixed value of the hopping strength ($J=0.08 U$) 
for several values of the dissipation strength $\Gamma_{\rm PA}$.
The initial state at $t=0$ is chosen to be the unit filling Mott insulating state $|\psi_{\rm Mott} \rangle$, 
i.e., the ground state of $\hat{H}_{\rm BH}$ in Eq.~(\ref{eq:bhh}) with $J=0$ and no dissipation. 
Note that this setup was realized in the quantum simulation experiment by means of sudden changes of the optical lattice depth and the intensity of the PA laser~\cite{tomita2017observation}.
We assume that the total lattice number is $N =L^3 = 10^3 = 1000$ sites forming a cubic lattice with periodic boundary conditions.
The measurable quantity to characterize the stability is the expectation values of the onsite atomic density $n_{j}(t) = {\rm Tr}\left( {\hat \rho}_{\rm eff}(t) {\hat n}_{j} \right)$ with ${\hat n}_{j} =\hat a_j^\dag \hat a_j$.
If the PA laser is turned off, the density does not change in time because of the number conservation law of the system associated with the global U(1) symmetry. 
However, the number conservation law is violated due to the two-body losses driven by the PA laser.
Figures~\ref{fig:loss_rate}(a) and \ref{fig:loss_rate}(b) display the results calculated by the SU(3) dTWA 
and the SU(3) FP-TWA, respectively.
As shown in these figures, the expectation value of the atomic density $n_j(t)$ monotonically decreases in time for all three typical values of the dissipation 
strength $\Gamma_{\rm PA}$ represented here by a dimensionless parameter $\gamma = \hbar \Gamma_{\rm PA}/U$.

In order to extract the two-body loss rate $\kappa$ from the behavior of the decay in $n_j(t)$, we utilize a fitting function of the form 
$f_{\kappa} (t) = (1+\kappa t)^{-1}$ for the early time regimes satisfying $n_{j}(t) > 0.4$~\cite{tomita2017observation}.
Note that this fitting function is the solution of a damping equation of the density associated with two-body losses, i.e., $\frac{dn_j}{dt} = - \kappa n^2_j$ with the initial conditions $n_{j}(t=0) = 1$~\cite{syassen2008strong}.
Figure~\ref{fig:loss_rate}(c) shows the loss rate $\kappa$ for the SU(3) dTWA as a function of $\gamma$. 
Note that $\gamma$ is independent of the optical lattice depth.
Here we assume the recoil energy of the optical lattice $E_{\rm R}/\hbar \approx 2\pi \times 4.0$ kHz for the lattice constant 
$d_{\rm lat}=266$ nm.
One of the main findings in Fig.~\ref{fig:loss_rate}(c) is the manifestation of the continuous quantum Zeno effect in the vicinity of 
$\gamma \sim 1$, which is qualitatively consistent with the experimental observation in Ref.~\cite{tomita2017observation}.
For more quantitative comparison, we also show the corresponding experimental results in the same figure and find 
that they are in good agreement with our theoretical results within the experimental error bars.
In the weak dissipation regime with $\gamma \lesssim 1$, the loss rate $\kappa$ becomes larger with increasing $\gamma$.
In contrast, in the strong dissipation regime with $\gamma \gtrsim 1$, the loss rate $\kappa$ decreases 
with increasing $\gamma$, which can be attributed to the suppression of doubly occupied sites due to strong two-body losses, 
i.e., the Zeno effect~\cite{tomita2017observation}.

Moreover, the loss rate estimated within the SU(3) dTWA is much better than the mean-field result 
$\hbar \kappa_{\rm MF}=16 z \gamma J^2 U^{-1} /(4+\gamma^2) \equiv 4z \hbar \Gamma_{\rm eff}$ from a previous  
work~\cite{garcia2009dissipation}, where $z=2d = 6$ is the coordination number of the lattice geometry.
This mean-field result is obtained by performing a site-decoupling approximation for the dynamical equation of 
$\langle {\hat n}_{j} \rangle$ with mapping to a hard-core boson model from the master equation in Eq.~(\ref{eq: lindblad}). 
The derivation of this effective model is based on a second-order expansion of the hopping term to adiabatically integrate Fock states, 
including doubly occupied sites, which is validated under the conditions of strong dissipation and strong interaction, i.e., $J \ll \hbar\Gamma_{\rm PA}$ and $J \ll U$.
As shown in Fig.~\ref{fig:loss_rate}(c), the mean-field treatment also captures the Zeno effect, and the curve of $\kappa_{\rm MF}$ exhibits a maximum at $\gamma = 2$, which is comparable to the dTWA result as well as the experimental result.
In particular, the dTWA result and the mean-field result are in good agreement for $\gamma \lesssim 1$.
However, the mean-field result overestimates the loss rate for $\gamma \gg 1$.
The significant deviations in the strong dissipative regime imply that the mean-field approximation for the hard-core boson representation ignores important effects that suppress loss events more effectively, which are successfully captured in the dTWA.

\begin{figure*}[!t]
\includegraphics[width=180mm]{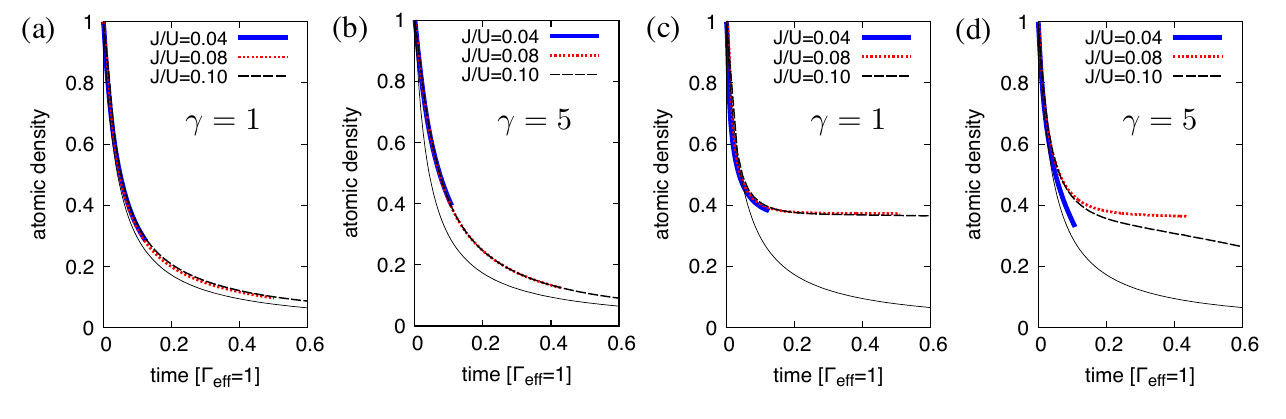}
\vspace{-10mm}
\caption{
Rescaled time evolution of the atomic density in units of the effective frequency $\Gamma_{\rm eff}$.
(a) The SU(3) dTWA results for $\gamma = 1$.
The blue solid, red dotted, and black dashed lines correspond to $J/U=0.04$, $0.08$, and $0.1$, respectively.
Each line is calculated until $t = 100 \hbar/U$.
For comparison, the mean-field result $n_{\rm MF}(t)=1/(1+4z \Gamma_{\rm eff} t)$ is also plotted by a black solid line.
(b) Same as (a) but for $\gamma = 5$. 
(c) Same as (a) but for the SU(3) FP-TWA results. 
(d) Same as (c) but for $\gamma = 5$.
}
\label{fig:scale}
\end{figure*}

The effective frequency $\Gamma_{\rm eff} = 4J^2\Gamma_{\rm PA}/(4U^2 + \hbar^2\Gamma_{\rm PA}^2)$ gives a natural time scale 
for a configuration to decay in the hard-core boson limit~\cite{garcia2009dissipation}.
Indeed, starting from a configuration with two neighboring particles, there is a process dominating over the loss, 
where a particle hops to the neighboring site with the amplitude $J/\hbar$, and then 
the particles decay at a doubly occupied site 
with the rate $\Gamma_{\rm PA}$.
Up to the second order with respect to $J / \hbar\Gamma_{\rm PA}$ and $J / U$ in terms of hard-core bosons, 
it gives rise to direct decay from the configuration to the vacuum with the rate $\Gamma_{\rm eff}$.
Moreover, in the hard-core boson description, the decay of the expectation value of the total particle number, and therefore 
those of the local atomic density starting from a homogeneous initial state as well, is solely determined by $\Gamma_{\rm eff}$~\cite{garcia2009dissipation}.
Rescaling the timescale of the loss dynamics by $\Gamma_{\rm eff}$, we find that all the results of the atomic density fall into 
a single function, independently of the values of $J/U$, as demonstrated in Fig.~\ref{fig:scale}.
This universal scaling behavior clarifies a nontrivial feature that the SU(3) dTWA captures well the two-step processes of decay, 
which are quantum mechanical, although the nonzero hopping term is only approximately treated as classical nonlinear interactions 
in this approach. 
However, the SU(3) FP-TWA fails to satisfy the scaling behavior, especially for strong dissipation, as shown in Fig.~\ref{fig:scale}(d).

We note that in the experimental setup of Ref.~\cite{tomita2017observation}, the controllable range of $\gamma$ is limited to 
$ 0 \leq \gamma \lesssim 5$.
Our SU(3) dTWA results strongly support that the continuous Zeno effect is significantly enhanced when $\gamma$ is further raised 
above $\gamma = 5$, beyond the experimentally accessible range.
This is in accordance with the intuition anticipated for the Zeno effect and makes the experimental observation of the Zeno effect in Ref.~\cite{tomita2017observation} more substantial. 
Additionally, as shown in Fig.~\ref{fig:loss_rate}(d), the SU(3) FP-TWA also produces a similar curve of $\kappa$ 
as a function of $\gamma$, 
implying almost the same threshold to the Zeno regime, but it has typically larger error bars in estimating the values 
due to the nonlinear properties of the Langevin equation.
Within the error bars, the values of $\kappa$ are comparable to those obtained by the SU(3) dTWA, 
and therefore the SU(3) FP-TWA result is also better than the mean field result for the estimation of the loss rate.

\begin{figure}[!t]
\begin{center}
\includegraphics[width=85mm]{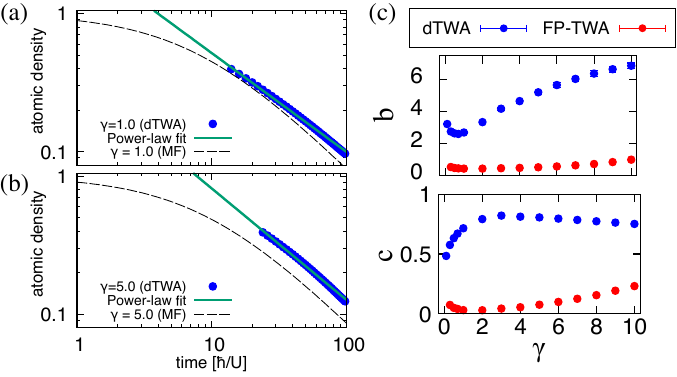}
\vspace{0mm}
\caption{
(a) 
Power-law fitting (green line) of the long-time behavior of the atomic density $n_j(t)$ obtained from the SU(3) dTWA 
(blue circles) for $J = 0.08 U$ and $\gamma = 1$. 
For comparison, the mean-field result is also shown by the dashed line. 
The power-law function used for fitting is $f_{\rm power}(t) = b (tU/\hbar)^{-c}$, where $b$ and $c$ are fitting parameters. 
Note that both the vertical and horizontal axes are in logarithmic scale. 
(b) 
Same as (a), but for $\gamma = 5$.
(c) Fitting parameters $b$ and $c$ in the fitting function $f_{\rm power}(t) = b (tU/\hbar)^{-c}$ for $J=0.08U$ with varying values 
of $\gamma$, obtained from the results evaluated by the SU(3) dTWA (blue circles) and the SU(3) FP-TWA (red circles). 
}
\label{fig:logntimefit}
\end{center}
\end{figure}

Let us further discuss the qualitative difference between these two semiclassical descriptions by comparing the results in more detail.
The onsite linearity of the SU(3) dTWA leads to long-time behavior for $tU/\hbar \gtrsim 10$ that is qualitatively distinguishable 
from that in the SU(3) FP-TWA.
In particular, for the typical values of $\gamma$ in Fig.~\ref{fig:loss_rate}, the density $n_j(t)$ at, for example, $t U/\hbar = 100$ 
in the SU(3) dTWA significantly deviates from that in the SU(3) FP-TWA, even though the short-time behavior agrees well 
for $tU/\hbar \lesssim 10$.
Moreover, as shown in Fig.~\ref{fig:loss_rate}(b), $n_j(t)$ obtained by the SU(3) FP-TWA for $\gamma = 1$ and $5$ crosses 
at a certain time, while there is no such intersection found for the SU(3) dTWA results in Fig.~\ref{fig:loss_rate}(a). 
Furthermore, in Figs.~\ref{fig:logntimefit}(a) and \ref{fig:logntimefit}(b), we observe that the long-time behavior of the density $n_j(t)$ 
predicted by the SU(3) dTWA can be well 
fitted with a power-law decay function with an exponent $c < 1$, implying that $n_{j}(t) \approx b (tU/\hbar)^{-c}$ for $t U/\hbar \gg 1$.
This power-law behavior differs from the mean-field result, indicating that the SU(3) dTWA captures nontrivial effects beyond 
those described by the simple mean-field approach.  
While the SU(3) FP-TWA results can also be fitted using the same power-law function, 
the exponents differ significantly, as shown in Fig.~\ref{fig:logntimefit}(c).

To gain insight into the validity of the TWA descriptions, we will compare the semiclassical results for a small cluster 
with the numerically exact results of the master equation in Sec.~\ref{exact}. 
There, we find that the long-time behavior obtained by the SU(3) dTWA is qualitatively similar to the exact results, 
and that the SU(3) FP-TWA is less qualitative than the SU(3) dTWA due to the nonlinearity of the dissipative contribution. 
Therefore, we conclude that the SU(3) dTWA results in Fig.~\ref{fig:loss_rate} are more reliable.

\subsection{Linear sweep time sequence of the hopping}
\label{subsec:sweep}

\begin{figure*}
\includegraphics[width=180mm]{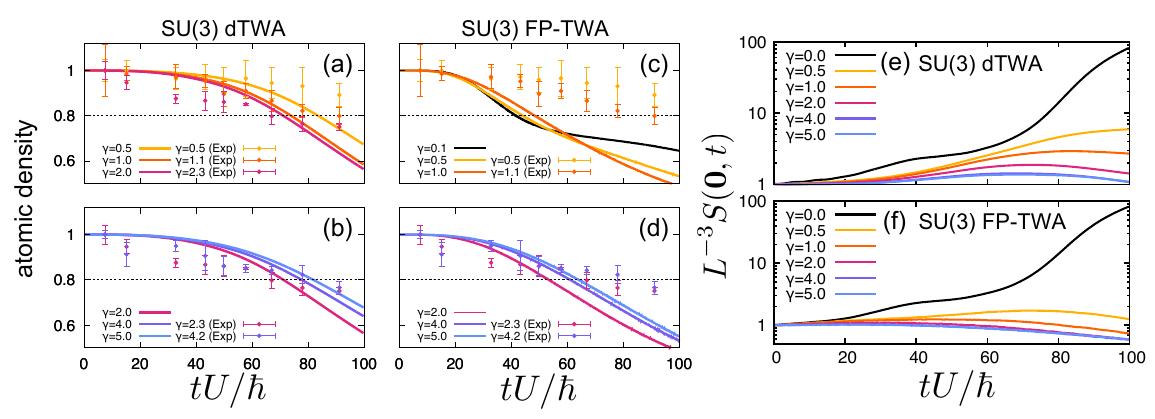}
\vspace{-7mm}
\caption{
(a), (b) 
SU(3) dTWA results of the nonunitary time evolution of the atomic density $n_j(t)$ for the atom loss dynamics, 
started from the unit filling Mott-insulating state $|\psi_{\rm Mott}\rangle$ at time $t=0$, 
with the linear sweep time sequence crossing the ground-state phase transition point of the Bose-Hubbard Hamiltonian, i.e., 
$J_{\rm ini}=0$ and $J_{\rm fin}=0.04 U$. The values of the dissipation strength $\gamma$ are indicated in the figures. 
The time grid size for the Runge-Kutta scheme is set to be $\Delta t = 2.0 \times 10^{-3} \hbar/U$.
(c), (d)
Same as (a) and (b) but for SU(3) FP-TWA results. Here, 
the time grid size for the Euler-Maruyama scheme is set to be $\Delta t = 2.0 \times 10^{-4} \hbar/U$. 
For comparison, the experimental results in Ref.~\cite{tomita2017observation} are also plotted by circles with error bars in (a)--(d). 
(e), (f) 
Time evolution of the phase coherence extracted from the single-particle correlation functions, i.e., $S({\bm 0},t)$, 
for the linear sweep time sequence evaluated by 
(e) the SU(3) dTWA and (f) the SU(3) FP-TWA.
}
\label{fig:melting}
\end{figure*}

Next, we analyze how the two-body loss dissipation affects the coherent time evolution caused by a slow parameter change 
in the Hamiltonian. 
At zero temperature and for a commensurate mean filling, the Bose-Hubbard Hamiltonian $\hat{H}_{\rm BH}$
exhibits a second-order quantum phase transition between a Mott-insulating state and a superfluid state~\cite{fisher1989boson,capogrosso2007phase}. 
In the three-dimensional setup with the mean occupation tuned to unit filling, the phase transition occurs at $J_{\rm c} = 0.0341 U$.
This exact critical value was numerically estimated in Ref.~\cite{capogrosso2007phase} by employing a quantum Monte Carlo method.
Note that the mean-field Gutzwiller approximation yields the critical point $J_{\rm c}^{\rm MFA} = 0.0286 U$~\cite{oosten2001quantum}, which is smaller than the exact value due to the less accuracy of the site-decoupling approximation.

Here, in our simulations, we prepare the unit filling Mott-insulating state $|\psi_{\rm Mott}\rangle$ 
at $t=0$ and then slowly drive it with the Lindblad equation for a time-dependent hopping amplitude at a fixed value of 
the dimensionless dissipation parameter $\gamma$. 
To realize a setup that is close to the real experiment, we assume a linear sweep time sequence of the hopping given 
by $J(t) = J_{\rm ini} + (J_{\rm fin}-J_{\rm ini})(t/ \tau)$ for $ 0 \leq t \leq \tau$~\cite{tomita2017observation}.
Here, $J_{\rm ini}$ and $J_{\rm fin}$ are the initial and final hopping amplitudes, which are set in the Mott-insulating and superfluid 
regimes, respectively.
During this sweep, the onsite interaction strength $U$ remains constant. 
In the experiment reported in Ref.~\cite{tomita2017observation}, the tuned parameter is the optical lattice depth, 
which is linearly ramped down from the Mott-insulating regime to the superfluid regime.
The dominant effect arising in the dynamics during the ramping down is the exponential increase of the hopping energy.
Hence, the linear sweep time sequence of the hopping provides a reasonable simplification of the time sequence realized 
in the experiment.

Figures~\ref{fig:melting}(a) and \ref{fig:melting}(b) show the SU(3) dTWA results of the atomic density $n_j(t)$ for the atom loss 
dynamics with the time sequence of the hopping amplitude $J(t)$ described above. 
We specifically choose $J_{\rm ini} = 0$ and $J_{\rm fin} = 0.04 U$ such that 
$J_{\rm ini} <  J_{\rm c}^{\rm MFA} < J_{\rm c} < J_{\rm fin}$ is satisfied.
The sweep time is taken to be $\tau = 100 \hbar/U$, for which non-adiabatic excitations of the system are sufficiently suppressed during the sweep. 
For the weak dissipation strength such as $\gamma=0.5$, the loss of atoms becomes significant at an onset time $t U / \hbar \sim 83.5$.
Here, the onset time is defined as the time when the atomic density drops below $0.8$, as indicated by a dotted line in Fig.~\ref{fig:melting}(a).
As shown in Fig.~\ref{fig:melting}(a), the onset time 
shifts to an earlier time with increasing $\gamma$ for $\gamma \lesssim 2$.
Indeed, the onset times for $\gamma = 1$ and $\gamma = 2$ are $t U / \hbar \sim 73.9$ and $t U / \hbar \sim 71.1$, respectively.
However, for the strong dissipation strength, i.e., $\gamma \gtrsim 2$, the onset time shifts to a later time with 
increasing $\gamma$, as shown in Fig.~\ref{fig:melting}(b). 
This result manifests the continuous quantum Zeno effect in the sweep dynamics, and a similar behavior has also been observed 
experimentally in Ref.~\cite{tomita2017observation}. 
The threshold strength of $\gamma$ separating the Zeno and non-Zeno regimes is $\gamma \sim 2$, 
which also agrees well quantitatively with the experimental observation~\cite{tomita2017observation}.
Since the numerically exact result for a small cluster also supports a similar onset value, 
as demonstrated in Sec.~\ref{exact}, our result here is considered to be qualitatively reliable.

The SU(3) FP-TWA also describes the delay of the atomic loss in the sweep dynamics for the strong dissipation strength. 
However, it gives rise to a qualitatively different result, as shown in Figs.~\ref{fig:melting}(c) and \ref{fig:melting}(d).
Namely, in contrast to the dTWA result, the delay occurs already for $\gamma \gtrsim 0.1$, implying that the Zeno regime 
is broadened on the side of smaller $\gamma$. 
For instance, along the time slice at $t = 40 \hbar/U$ in Figs.~\ref{fig:melting}(c) and \ref{fig:melting}(d), the atomic density is found 
to be significantly suppressed even for $\gamma = 0.1$, and it starts to increase with increasing $\gamma$ beyond the threshold. 
In other words, the nonlinearity of the FP-type approximation manifests itself in an excessive suppression of the atomic density, 
thus broadening the Zeno regime in the SU(3) FP-TWA.

In Figs.~\ref{fig:melting}(a)--\ref{fig:melting}(d), we also plot the experimental results of the atomic density 
for $\gamma = 0.5$, $1.1$, $2.3$, and $4.2$ reported in Ref.~\cite{tomita2017observation}, which should be directly compared with 
our TWA results for $\gamma = 0.5$, $1$, $2$, and $4$, respectively. 
Here, the experimental results are plotted 
along the time axis such that the values of $J/U$ during the sweep in the experiment are the same as the corresponding values 
in the linear sweep time sequence employed in our simulations. 
As shown in these figures, we find that, specifically in the SU(3) dTWA, the qualitative behavior of the $\gamma$ dependence 
of the melting dynamics is reasonably captured within the experimental error bars.

We further analyze the suppression of the superfluid phase coherence in the presence of the strong two-body loss dissipation. 
For this purpose, 
we consider a spatial Fourier transform of the equal-time single-particle correlation function, i.e., the momentum distribution, 
defined by 
\begin{align}
S({\bf k},t) = \sum_{{\bf R}_i,{\bf R}_j} e^{i{\bf k}\cdot ({\bf R}_i-{\bf R}_j)} 
{\rm Tr} \left[ \hat{\rho}_{\rm eff}(t) {\hat a}^{\dagger}_i {\hat a}_j \right],
\end{align}
where ${\bf R}_i$ represents the real-space vector specifying site $i$ and 
${\bf k}$ denotes the momentum vector in three dimensions.
Note that this distribution is measurable in experiments by using the time-of-flight (TOF) imaging techniques~\cite{lewenstein2012ultracold}.
Here, we specifically set ${\bf k}=(0,0,0)$ for the purpose of quantifying the emergence of coherence over long distances after the hopping sweep.
In addition, we note that no inhomogeneity effect due to a trapping potential for the gas is taken into account here.

Figure~\ref{fig:melting}(e) shows the SU(3) dTWA results for $S({\bf 0},t)$.
In the absence of dissipation, i.e., $\gamma = 0$, the growth of long-distance correlations is recovered within this approximation, 
implying a dynamical phase transition of the system from the Mott-insulating phase to a superfluid phase.
For the weak dissipation strength, i.e., $\gamma \leq 1$, this semiclassical method remains to produce large occupation of the zero-momentum distribution at the final state, i.e., $L^{-3}S({\bf 0},t=\tau) \gg 1$.
Hence, the phase coherence comprised of coherent bosons survives well against the weak dissipation.
However, for the strong dissipation strength, i.e., $\gamma \gtrsim 2$, the distribution is found to be strongly suppressed 
such that $L^{-3}S({\bf 0},t=\tau) \sim 1$, 
implying that the strong dissipation stabilizes the Mott-insulating state and leads to the delay of the evolution toward a superfluid phase due to the continuous Zeno effect. 
This result is consistent with the experimental observation in TOF absorption images~\cite{tomita2017observation}, revealing 
that the phase coherence associated with superfluidity is suppressed by the strong dissipation.

On the other hand, the SU(3) FP-TWA gives rise to much stronger suppression of the phase coherence because of its nonlinearity 
in the dissipation force.
For example, as shown in Fig.~\ref{fig:melting}(f), $L^{-3}S({\bf 0},t=\tau) \sim 1$ is realized even for the weak dissipation strength 
such as $\gamma = 0.5$.
Moreover, for $\gamma \gtrsim 1$, the distribution at the final state is significantly suppressed, i.e., 
$L^{-3}S({\bf 0},t=\tau) \ll 1$, also implying the loss of phase coherence.

\subsection{Restored superfluid coherence after turning off the PA light}

\begin{figure}
\includegraphics[width=84mm]{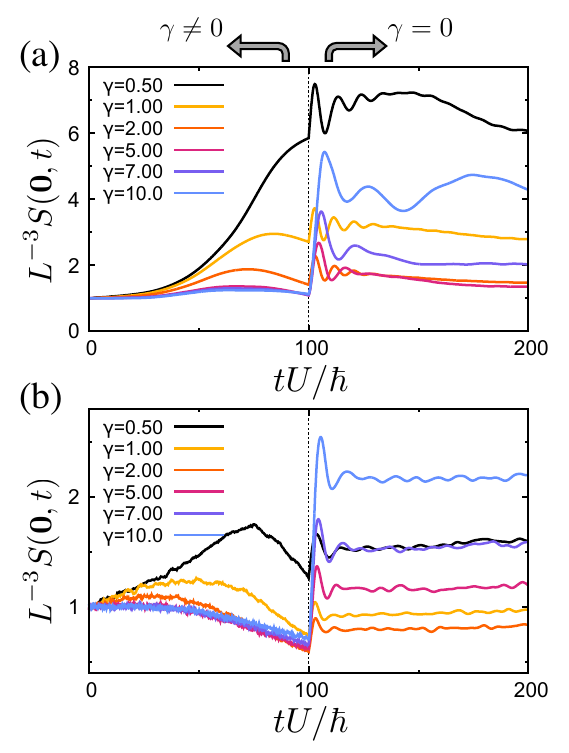}
\vspace{-5mm}
\caption{
(a) SU(3) dTWA results for the nonunitary time evolution of the zero momentum distribution $S({\bf k}={\bf 0},t)$, 
where $\gamma$ varies from $\gamma\ne0$ (indicated in the figure) to $\gamma=0$ at $t=\tau$, 
mimicking the abrupt turning off of the PA light at $t = \tau$.
For $t\le\tau$, the linear sweep time sequence is applied, as in Fig.~\ref{fig:melting}, while 
the hopping $J(t)=J_{\rm fin}$ for $t>\tau$. 
The vertical dotted line indicates the endpoint of the sweep at $t=\tau$.
(b) Same as (a) but for the SU(3) FP-TWA results. 
}
\label{fig:turnoff_PA}
\end{figure}

\begin{figure}
\begin{center}
\includegraphics[width=80mm]{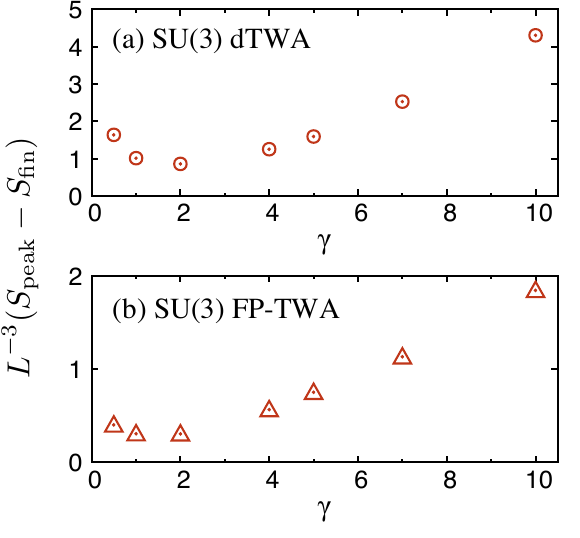}
\vspace{-4mm}
\caption{
The restored peak height $\delta h = L^{-3} ( S_{\rm peak}-S_{\rm fin} )$ after turning off the PA laser, i.e., by setting 
$\gamma=0$ at the end point of the sweep (see Fig.~\ref{fig:turnoff_PA}), 
vs the dissipation strength $\gamma$.
The results are evaluated by (a) the SU(3) dTWA and (b) the SU(3) FP-TWA.
}
\label{fig:peaks}
\end{center}
\end{figure}

Finally, we examine whether the SU(3) TWA  can capture the restoration of the superfluid coherence after suddenly turning 
off the PA laser at $t = \tau$.
A similar setup was implemented for the lattice ramp-down experiment in Ref.~\cite{tomita2017observation}.
It was observed in the experiment that after a certain hold time of the gas, an interference pattern in TOF images is restored, 
implying the phase coherence reformation~\cite{tomita2017observation}.
This suggests that the suppression of superfluid coherence is not attributed to the heating caused by the PA laser 
and that the final state after the ramping down remains a Mott-insulating state.

Figure~\ref{fig:turnoff_PA} shows the SU(3) dTWA and FP-TWA results simulating the time evolution when $\gamma$ 
abruptly changes from $\gamma \neq 0$ to zero at $t = \tau$. 
We find that the increase of the zero-momentum distribution occurs for all $\gamma$ studied after turning off the dissipation,
implying the restoration of the phase coherence over long distances, as observed experimentally in 
the visibility measurements of TOF images~\cite{tomita2017observation}. 
In the time region where $\gamma=0$, i.e., $t>\tau$, 
the momentum distribution at ${\bf k}={\bf 0}$ first increases up to a maximum peak, and then eventually saturates. 
Within the SU(3) TWA, the long-time evolution for $t \gg \tau$ is expected to lead to a classical thermal state described by the Gibbs 
distribution for the classical chaotic Hamiltonian~\cite{wurtz2018cluster}, i.e., the SU(3) representation of the Bose-Hubbard model 
with a conserved filling factor.

In order to better characterize the properties of the restored peaks in the momentum distribution, let us define 
$S_{\rm peak} = \max_{t>\tau}[ S({\bf k}={\bf 0},t) ]$ and $S_{\rm fin} = S({\bf k}={\bf 0},t=\tau)$. 
Figure \ref{fig:peaks} displays the restored peak height $\delta h = L^{-3} ( S_{\rm peak}-S_{\rm fin} )$ as a function of $\gamma$. 
In the SU(3) dTWA results shown in Fig.~\ref{fig:peaks}(a), the peak height $\delta h$ increases with increasing $\gamma$ for 
$\gamma \gtrsim 2$, while it decreases for $\gamma \lesssim 2$.
The increase of $\delta h$ for $\gamma \gtrsim 2$ implies that many more particles stay in the system due to the stronger dissipation, 
i.e., the Zeno effect, before switching off the dissipation $\gamma$.
Given the results in Figs.~\ref{fig:melting}(a) and \ref{fig:melting}(b) on the atomic density in the linear sweep time sequence, 
the restored peak height $\delta h$ turns out to follow the total number of particles surviving in the optical lattice at the end point of the sweep, i.e., $t=\tau$.

As shown in Fig.~\ref{fig:peaks}(b), a similar behavior of the $\gamma$ dependence of the peak height 
is also found in the SU(3) FP-TWA. 
Namely, $\delta h$ first decreases with increasing $\gamma$ and then starts to increase at 
a minimum point located around $1 \leq \gamma \leq 2$.
The presence of the minimum in this regime is also correlated to the behavior of the final state atom densities at $t=\tau$ 
shown in Figs.~\ref{fig:melting}(c) and \ref{fig:melting}(d).
However, we should note that the final states at $t=\tau$ are not reliable, even qualitatively, in the SU(3) FP-TWA, 
as implied by the numerically exact results in Sec.~\ref{exact}, and therefore we consult the SU(3) dTWA results 
to be compared with the experiment.

\subsection{Numerically exact results}
\label{exact}

\begin{figure}
\includegraphics[width=80mm]{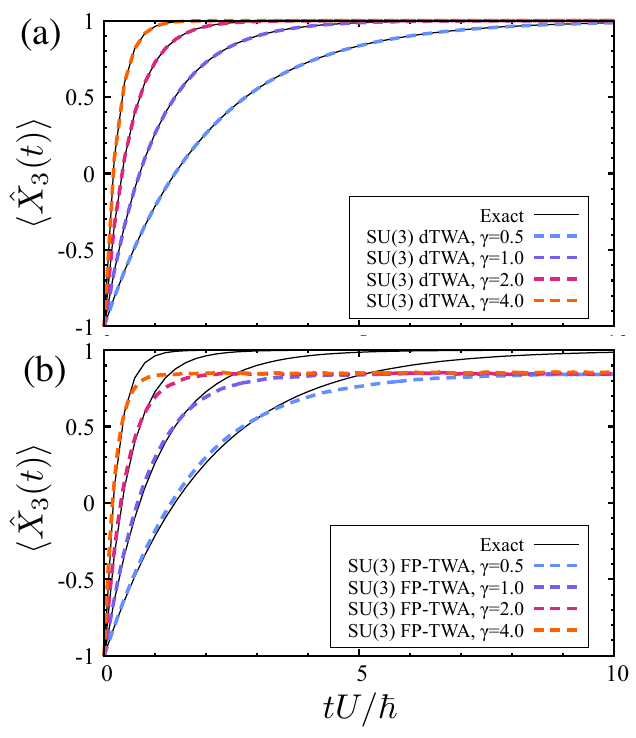}
\vspace{-0mm}
\caption{
(a) Time evolution of $\langle X_3(t)\rangle$ for the single-site system with the two-body loss, corresponding to the $J=0$ limit 
of the dissipative Bose-Hubbard model, evaluated by the SU(3) dTWA.
The initial state at $t=0$ is chosen to be the doubly occupied state $|2\rangle$.
This state is the simultaneous eigenstate of ${\hat X}_3$ and ${\hat X}_8$, i.e., ${\hat X}_3|2\rangle = -|2\rangle$ 
and ${\hat X}_8|2\rangle = -\frac{1}{\sqrt{3}}|2\rangle$. 
Here, 1000 trajectories are used to obtain the smooth results as a function of time $t$. 
(b) Same as (a) but for the SU(3) FP-TWA results. Here, 100\;000 trajectories are required to obtain the smooth results. 
For comparison, the exact solution of the master equation for each value of $\gamma$ is also provided by the solid line.
}
\label{fig:singlesite}
\end{figure}

\begin{figure*}
\includegraphics[width=182mm]{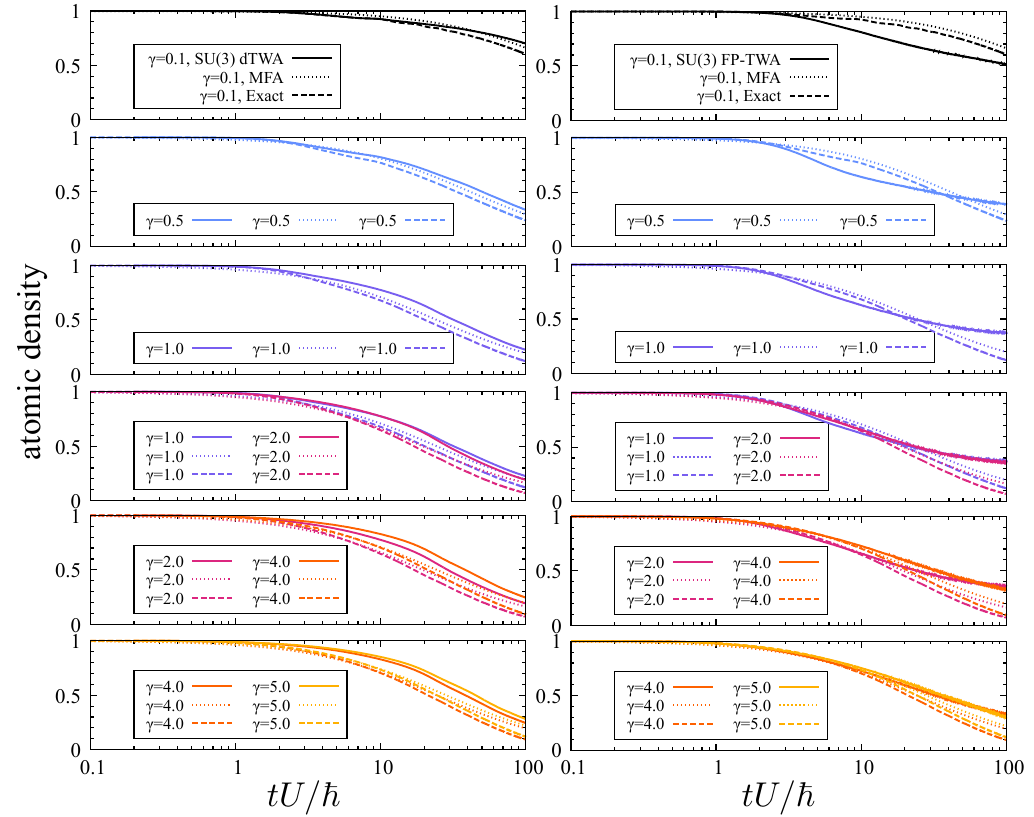}
\vspace{-7mm}
\caption{
SU(3) dTWA (left panels) and FP-TWA (right panels) results of the nonunitary time evolution of the atomic density $n_j(t)$ for 
the dissipative Bose-Hubbard model on a four-site ring (solid lines). 
The initial state at $t=0$ for the time evolution is the unit filling Mott-insulating state $|\psi_{\rm Mott}\rangle$. 
The hopping amplitude $J$ is the same as in Fig.~\ref{fig:loss_rate}, i.e., $J = 0.08 U$, and the values of the dissipation strength 
$\gamma$ are indicated in the figures. 
For comparison, the numerically exact and mean-field results are also plotted by dashed and dotted lines, respectively. 
In the SU(3) dTWA, the time grid size for the explicit fourth-order Runge-Kutta scheme is set to 
$\Delta t = 2.0 \times 10^{-2} \hbar/U$ and 
the number of generated classical trajectories is $20\;000$.
In the SU(3) FP-TWA, the grid size for the first-order Euler-Maruyama scheme is set to 
$\Delta t = 2.0 \times 10^{-4}  \hbar/U $ and 
the number of generated Langevin trajectories is $200\;000$.
}
\label{fig:lossrate_exact}
\end{figure*}

\begin{figure*}
\includegraphics[width=170mm]{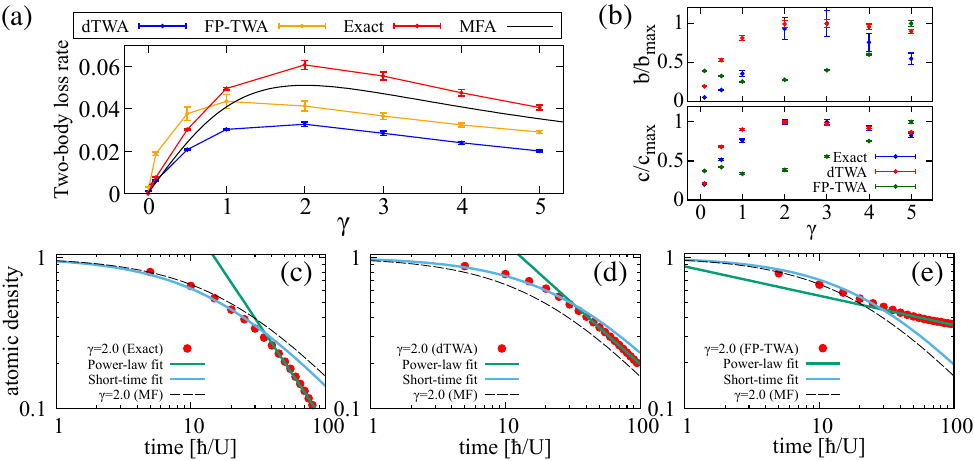}
\vspace{0mm}
\caption{
(a)
Two-body loss rate $\kappa$ as a function of $\gamma$ 
extracted from the nonunitary dynamics of the four-site system in Fig.~\ref{fig:lossrate_exact}.
The time unit for the loss rate is $\hbar/U$ and the data points of $n_j(t)$ used for fitting in the form 
$f_{\kappa}(t) =(1+\kappa t)^{-1}$ 
are separated by $5 \hbar/U$.
The exact (red) and SU(3) dTWA (blue) results show a maximum at $\gamma \sim 2$, while 
the SU(3) FP-TWA results (orange) exhibit a maximum at $\gamma \sim 1$ within the standard fitting errors.
The mean-field result (black) also shows a maximum at $\gamma = 2$.
To capture the early-time behavior, the fitting is performed on the corresponding data $n_j(t)$ that satisfy $t \leq 40 \hbar/U$.
(b) 
Power-law exponents for the long-time behavior of the atomic density $n_j(t)$ for varying values of $\gamma$.
The fitting range of time $t$ using the function $f_{\rm power}(t) = b(tU/\hbar)^{-c}$ is restricted to $tU/\hbar > 40$.
The values of $b$ and $c$ are rescaled by their corresponding maximum values, $b_{\rm max}$ and $c_{\rm max}$, where 
$(b_{\rm max},c_{\rm max}) = (39.6,1.3)$, $(8,0.8)$, and $(3.1,0.5)$ for the exact, SU(3) dTWA, and SU(3) FP-TWA 
results, respectively. 
(c) 
Power-law fitting using the function $f_{\rm power}(t)$ and short-time fitting using the function 
$f_{\kappa}(t)$ for the atomic density $n_j(t)$ obtained by the numerically exact calculations for $\gamma = 2$. 
For comparison, the mean-field result is also shown by a dashed line. 
(d)
Same as (c), but for the results obtained from the dTWA. 
(e)
Same as (c), but for the results obtained from the FP-TWA. 
}
\label{fig:fitparam_exact}
\end{figure*}

\begin{figure}
\includegraphics[width=80mm]{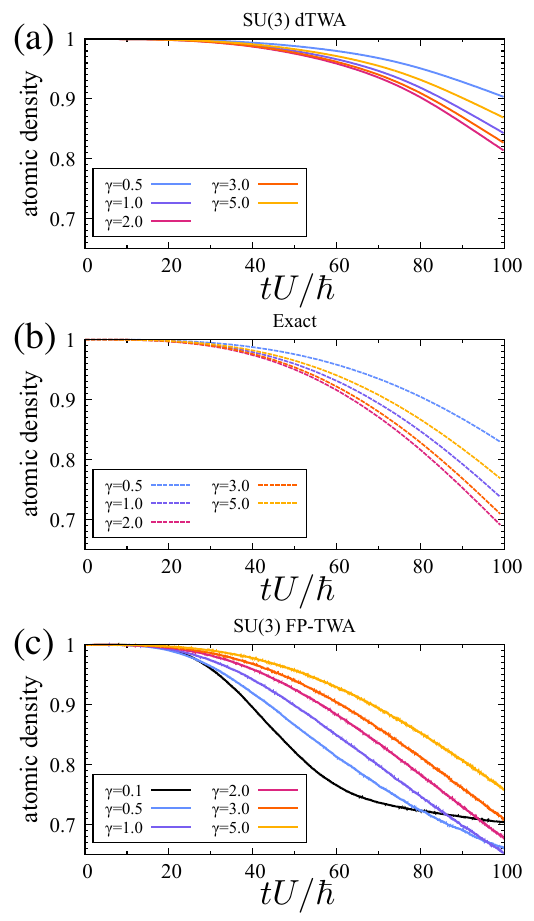}
\vspace{-3mm}
\caption{
(a) Same as Figs.~\ref{fig:melting}(a) and \ref{fig:melting}(b) but for a four-site ring. 
The number of generated classical trajectories is $20\;000$. 
(b) Same as (a) but for the numerically exact solutions. 
(c) Same as (a) but for the SU(3) FP-TWA results. 
The number of the generated Langevin trajectories is $200\;000$.
}
\label{fig:sweep_exact}
\end{figure}

Finally, we numerically examine the performance of the TWA schemes introduced here in this paper 
for the Lindblad master equation by comparing their results with the numerically exact ones.

When the hopping amplitude is zero, i.e., $J=0$, the SU(3) dTWA reproduces the exact quantum time evolution 
generated by the master equation with a direct product state for the typical initial state. 
This is attributed to the linearity of the classical equation of motion in the interaction $U$ term and the dissipation $\gamma$ term 
as implied in Eq.~(\ref{eq: dissipative_eom_dtwa}).
Figure \ref{fig:singlesite}(a) displays numerical results calculated by the SU(3) dTWA for a single-site system, 
demonstrating that this specific semiclassical representation successfully reproduces the exact solutions of the dynamics of 
$\langle {\hat X}_{3}(t) \rangle = {\rm Tr}[ \rho_{\rm eff}(t) T_3 ]$ for arbitrary dissipation strength at all time.

However, as shown in Fig.~\ref{fig:singlesite}(b), the quantitative validity of the SU(3) FP-TWA is verified only within a finite timescale. 
This limitation is due to the nonlinearity in the dissipation term of the Langevin equation.
For $\gamma = 0.5$, the SU(3) FP-TWA result coincides well with the numerically exact solution until $t U/\hbar \sim 3$.
After this timescale, the SU(3) FP-TWA result deviates from the numerically exact solution and finally converges to a false steady state 
with $\langle {\hat X}_{3}(t) \rangle < 1$ for $tU/\hbar \gg 1$.
Note that the exact steady state for the present setup is the trivial bosonic vacuum state $|0\rangle$, 
implying $\langle {\hat X}_{3}(t) \rangle = 1$ for $tU/\hbar \gg 1$.
For larger values of $\gamma$, i.e., $\gamma = 1$, $2$, and $4$, the threshold time is estimated as 
$tU/\hbar \sim 1.5$, $1$, and $0.5$, respectively. 
Hence, the valid timescale of this approximation shortens with increasing $\gamma$.
Notice also that the time-evolved states for these different values of $\gamma$ result in the same false steady state at long time.
Comparing the results in Figs.~\ref{fig:singlesite}(a) and ~\ref{fig:singlesite}(b), the local linearity of the dissipative force in the SU(3) 
dTWA is essential to reproduce the correct steady state reached via the successive onsite losses over time.

We further examine how well the semiclassical representations can reproduce the exact solutions of the master equation 
in a multiple-site case with nonzero hopping. 
For this purpose, let us consider a chain composed of $L=4$ sites under periodic boundary conditions.
Except for the spatial geometry of the system, the situation considered here is the same as that in Sec.~\ref{subsec:quench}. 
Namely, starting from the Mott-insulating state $|\psi_{\rm Mott}\rangle$ at unit filling, we suddenly turn on the dissipation $\gamma$ 
as well as the hopping $J$ at $t=0$ and numerically calculate the time evolution of the atomic density $n_j(t)$, where we choose 
$J=0.08U$.

The left panels of Fig.~\ref{fig:lossrate_exact} compare the SU(3) dTWA results of the atomic density $n_j(t)$ 
with the numerically exact solutions for different values of $\gamma$. 
For $\gamma = 0.1$, the result obtained by the SU(3) dTWA agrees very well with the exact solution until $t U/\hbar \sim 10$, but 
is merely qualitative at later time, where the deviation becomes larger as the time further evolves. 
However, for $\gamma \gtrsim 0.5$, the time regime in which the SU(3) dTWA description is quantitative is shortened 
up to $t U/\hbar \sim 1$. 
This is due to the approximate treatment of the offsite hopping term in the Bose-Hubbard Hamiltonian.
In other words, the SU(3) dTWA cannot quantitatively evaluate the dynamical creation of double occupation per site at long time, 
which occurs via tunneling of the atoms between sites. 
Regardless of such a limitation, it is noteworthy that the SU(3) dTWA can estimate a reasonable Zeno threshold value as 
compared to the exact dynamics, i.e., the delay of atom loss happens around $\gamma \sim 2$ in this one-dimensional system.
Note also that the SU(3) dTWA results are always larger than the corresponding exact solutions, implying that higher-order 
quantum corrections associated with offsite correlations give rise to the reduction of the atomic densities at all time.

Reflecting these consequences, we also find in Fig.~\ref{fig:fitparam_exact}(a) that the two-body loss rate $\kappa$ 
extracted from the dTWA results exhibits a maximum around at $\gamma \sim 2$, 
although the values themselves are smaller than the exact ones.
However, the agreement with the three-dimensional experiment in Fig.~\ref{fig:loss_rate}(c) suggests that the deviations between 
these two results of the loss rate may be smaller in higher dimensions because the valid timescale of the dTWA increases with 
the connectivity between sites, i.e., the coordination number $z=2d$~\cite{kunimi2021performance}.
Additionally, we perform the power-law fitting for the long-time behavior of the density in Fig.~\ref{fig:lossrate_exact}.
Figures~\ref{fig:fitparam_exact}(c)--\ref{fig:fitparam_exact}(e) also demonstrate that the atomic density $n_j(t)$, 
obtained from three different methods, fit well with the power-law function in the long-time regime, 
clearly distinguishing it from the short-time behavior, which is well fitted by $f_{\kappa}(t) = (1+\kappa t)^{-1}$. 
This qualitative difference is also evidence when compared to the mean-field results shown in the same figures. 
Notably, as shown in Fig.~\ref{fig:fitparam_exact}(b), the rescaled power exponent $c/c_{\rm max}$ 
in the  power-law function obtained from the SU(3) dTWA closely follow the exact results for varying $\gamma$, 
while the FP-TWA results deviate significantly.

The right panels of Fig.~\ref{fig:lossrate_exact} show the SU(3) FP-TWA results of the atomic density $n_j(t)$ 
compared with the numerically exact solutions.
For the weak dissipation strength, i.e., $\gamma = 0.1$, the SU(3) FP-TWA result deviates form the exact solution around 
$t U/\hbar = 2$.
However, for the strong dissipation strength such as $\gamma = 4$ and $5$, the SU(3) FP-TWA and exact results coincide 
rather well until $tU/\hbar \lesssim 6$.
We should note that this coincidence is not controlled in this approximation.
Although the SU(3) FP-TWA can capture the continuous Zeno effect for the strong dissipation strength, it estimates the threshold 
at $\gamma \sim 1$, which is smaller than the exact solution. 
As shown in Fig.~\ref{fig:fitparam_exact}(a), the maximum of the two-body loss rate $\kappa$ appears around the same value of 
$\gamma$.
Comparing the SU(3) dTWA and FP-TWA results, we conclude that the qualitative performance of the SU(3) FP-TWA is worse 
than that of the SU(3) dTWA at long time.

Finally, we evaluate the performance of these semiclassical methods in the case of the time-dependent linear sweep of the hopping 
amplitude discussed in Sec.~\ref{subsec:sweep}.
As shown in Fig.~\ref{fig:sweep_exact}(a), the onset of the Zeno effect estimated by the SU(3) dTWA 
is around $\gamma = 2$, which is consistent with the numerically exact result shown in Fig.~\ref{fig:sweep_exact}(b).
However, as shown in Fig.~\ref{fig:sweep_exact}(c), the FP-TWA estimates an incorrect onset at $\gamma \sim 0.1$.
Note that a similarly small onset of the Zeno effect is found in the three-dimensional SU(3) FP-TWA calculations 
in Fig.~\ref{fig:melting}(c). 
As already discussed in Fig.~\ref{fig:lossrate_exact}, we also note that higher-order corrections beyond the SU(3) dTWA reduce 
the atomic densities at all time for any $\gamma$. 
Finally, we should emphasize that although the qualitative behavior of the decay of the atomic density in the SU(3) dTWA is 
similar to that in the exact dynamics, apart from the systematic corrections with the same sign, the SU(3) FP-TWA fails to capture 
even such qualitative behavior.

\section{
Conclusions}
\label{sec:conclusions}

In conclusions, we have generalized the SU(${\cal N}$) TWA method for the Bose-Hubbard system to include the Lindblad dynamics 
in the presence of onsite dissipation terms.
We introduced two semiclassical representations of the dynamics governed by the effective Lindblad master equation: 
the discrete TWA representation and the Fokker-Planck TWA representation.
These methods were applied to the dissipative Bose-Hubbard system with the two-body losses, previously 
studied experimentally in Ref.~\cite{tomita2017observation}. 
We demonstrated that these semiclassical representations can qualitatively capture the essential features of 
the continuous quantum Zeno effect in the dissipative nonunitary dynamics, as observed experimentally. 
In particular, the discrete TWA representation was found to produce reasonable thresholds for the dissipation strength,  
distinguishing between the Zeno and non-Zeno regimes, which are comparable to the experimental results of the analog 
quantum simulation~\cite{tomita2017observation}.
This is attributed to the onsite linearity of the classical equation of motion in the discrete TWA representation, 
which was used in the Monte Carlo simulations.
The local linearity is highly powerful because a collection of fluctuating trajectories with locally linear equations results in a power-law 
decay in the long-time atomic density loss dynamics, a feature that cannot be captured by the naive mean-field approximation.

The general strategy that we have developed opens a new avenue for addressing a wide range of open quantum many-body 
systems, particularly in two and three dimensions, which are computationally challenging.
As a future study, it would be interesting to apply those SU(${\cal N}$) TWA methods to other experiments associated with 
dissipative Bose-Hubbard systems~\cite{labouvie2016bistability,bouganne2020anomalous} and discuss the reproducibility of nontrivial 
driven-dissipative dynamical phenomena reported therein. 
The generalization of the methods to SU(${\cal N}$) spin systems
for ${\cal N} > 3$
is straightforward, and these methods potentially enable one to 
evaluate the impact of dissipation on various experiments of analog quantum simulation, such as optical tweezer arrays loaded with 
Rydberg gases, trapped ions, and large-$S$ dipolar gases in optical lattices. 
Beyond the present capability of the methods based on onsite variables, it is also a promising direction to incorporate the ideas of 
the cluster TWA with fluctuating cluster variables~\cite{wurtz2018cluster} to treat offsite quantum correlations caused by the hopping 
term more accurately~\cite{Nagao2024}.

\begin{acknowledgments}

Parts of the numerical simulations were carried out on the HOKUSAI supercomputing system at RIKEN (Project ID No. Q22576) 
and the supercomputer Fugaku installed at the RIKEN Center for Computational Science. 
We thank Kazuhiro Seki for his helpful comments on MPI parallelization for Monte Carlo simulations.
We also thank Masaya Kunimi and Tetsuro Nikuni for fruitful discussions.
We acknowledge Takafumi Tomita and Yoshiro Takahashi for sharing the experimental data in Ref.~\cite{tomita2017observation}.
This work was financially supported by JSPS KAKENHI (Grants No. JP18H05228, No. JP21H01014, No. JP21H04446, 
and No. JP21H03455), by MEXT Q-LEAP (Grant No. JPMXS0118069021), by JST FOREST (Grant No. JPMJFR202T), and 
by JST COI-NEXT (Grant No. JPMJPF2221).
This work was also supported by the RIKEN TRIP initiative (RIKEN Quantum).

\end{acknowledgments}

\appendix

\section{
SU(2) semiclassical expressions of a dissipative Landau-Zener model
}
\label{LZ model}

To illustrate the applicability of our phase-space methods to other systems, in this appendix 
we analyze a simple system of a single Pauli spin 
coupled to a Markovian bath, which is described by the following Lindblad master equation: 
\begin{align}
\frac{d}{dt}{\hat \rho} = - i [{\hat H}_{\rm LZ},{\hat \rho}] + L_{\rm diss}[{\hat \rho}],
\end{align}
 where 
\begin{align}
{\hat H}_{\rm LZ} 
&= \frac{\omega}{2}{\hat \sigma}_{z} + \frac{\Delta}{2}{\hat \sigma}_{x}
\end{align}
is the Landau-Zener Hamiltonian and 
\begin{align}
L_{\rm diss}[{\hat \rho}] 
&= \frac{\gamma_0}{2}\left[2{\hat \sigma}_{-}{\hat \rho}{\hat \sigma}_{+} - {\hat \sigma}_{+}{\hat \sigma}_{-}{\hat \rho} -  {\hat \rho}{\hat \sigma}_{+}{\hat \sigma}_{-} \right] 
\end{align}
is the dissipative term for a decoherence process accompanying spin flips 
with $\hat\sigma_\pm=\hat\sigma_x \pm i\hat\sigma_y$. 
Here, $\hat\sigma_{x\,(y,\,z)}$ is the $x$ ($y,\,z$) component of 
the Pauli operator.

\begin{figure}[b]
\includegraphics[width=88mm]{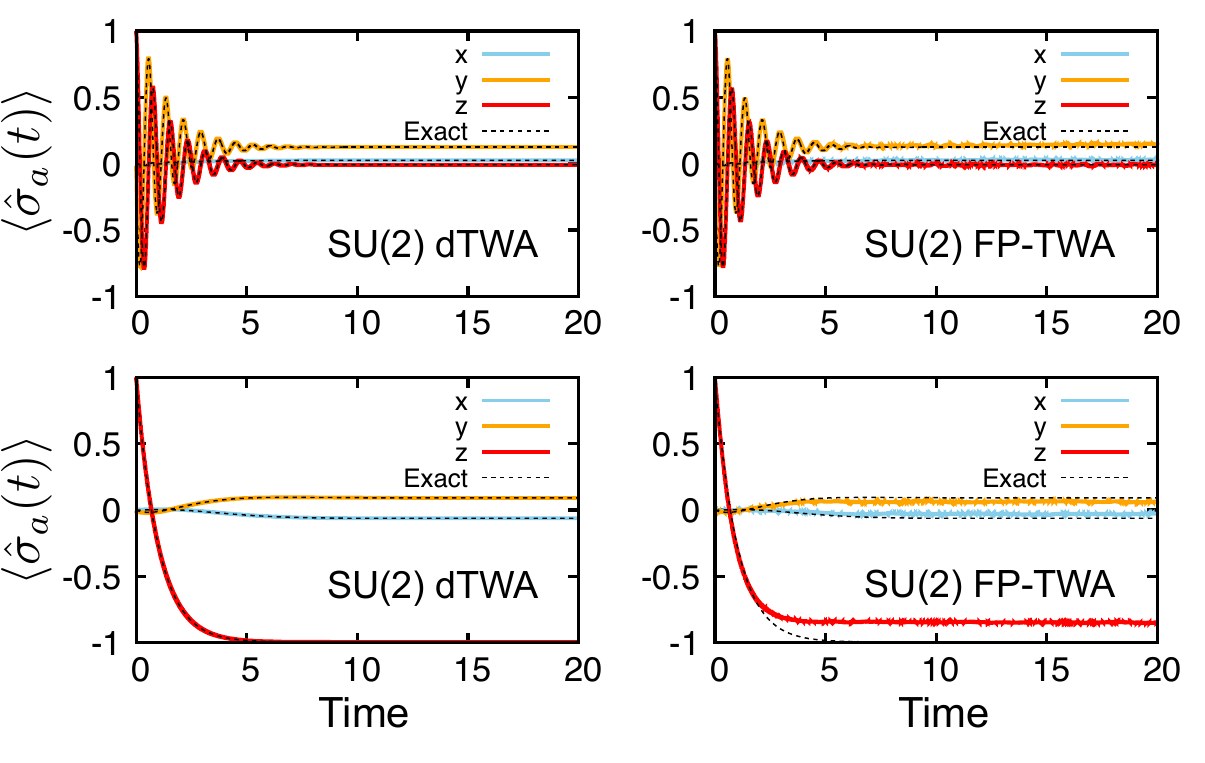}
\vspace{-6mm}
\caption{
Dynamics of the dissipative Landau-Zener model starting from the initial up-spin state 
${\hat \rho}(t=0)=| \!\! \uparrow\rangle\langle \uparrow \!\! |$ with $\hat\sigma_z| \!\! \uparrow\rangle=| \!\! \uparrow\rangle$.
The top panels represent the results for $(\omega,\Delta,\gamma_0)=(-0.1,8,1)$, while the bottom panels are for 
$(\omega,\Delta,\gamma_0)=(1/3,0.2/3,1)$.
The left (right) panels are calculated using the SU(2) dTWA (FP-TWA). 
The time unit for the horizontal axis is $\gamma_0 = 1$.
}
\label{fig:su2}
\end{figure}

This system is trivial because the dynamics from arbitrary initial states can be exactly solved within the SU(2) dTWA scheme. 
Indeed, the Hamiltonian ${\hat H}_{\rm LZ}$ is already linear in the Pauli operators, and the dissipative term $L_{\rm diss}[{\hat \rho}]$
can also be linearizable using the technique based on the phase-point operator. 
The left panels in Fig.~\ref{fig:su2} show the results demonstrating the solvability of the SU(2) dTWA 
for two typical sets of parameters $(\omega,\Delta,\gamma_0)=(1/3,0.2/3,1)$ and $(-0.1,8,1)$.

As for the SU(2) FP-TWA, making a similar approximation as for the SU(3) FP-TWA in Eq.~(\ref{eq: approx. bopp}), 
we obtain the Langevin equations for the classical spin: 
\begin{align}
dx 
&= f_x dt - \sqrt{\gamma_0}z \cdot dW_1, \\
dy 
&= f_y dt - \sqrt{\gamma_0}z \cdot dW_2, \\
dz
&= f_z dt + \sqrt{\gamma_0}(x \cdot dW_1 + y \cdot dW_2),
\end{align}
where the force vector ${\bm f}=(f_x,f_y,f_z)$ is derived as follows: 
\begin{align}
f_{x} 
&= \frac{\gamma_0}{2}xz - \frac{\gamma_0}{2}x - \omega_0 y, \\
f_{y} 
&= \frac{\gamma_0}{2}yz - \frac{\gamma_0}{2}y + \omega_0 x - \Delta z, \\
f_{z} 
&= - \frac{\gamma_0}{2}(x^2+y^2) - \gamma_0 z + \Delta y.
\end{align}
The stochastic noise terms $dW_1$ and $dW_2$ satisfy the normalization $dW_{1}^2 = dW_{2}^2 = dt$ and $dW_{1}dW_{2} = 0$. 
For $(\omega,\Delta,\gamma_0)=(-0.1,8,1)$, as presented in Fig.~\ref{fig:su2}, each spin component relaxes to 
a steady value as expected, which is well described even within the SU(2) FP-TWA.
The final state is sufficiently close to the infinite-temperature mixed state ${\hat \rho}_{t = \infty} =  2^{-1}{\hat I} + 2^{-1}\sum_{a=x,y,z} {\rm Tr}({\hat \sigma}_{a} {\hat \rho}_{t=\infty}) {\hat \sigma}_{a} \approx 2^{-1}{\hat I}$.
Hence, the system is far from the trivial vacuum state for the dissipative term, i.e., the down-spin state $|\!\! \downarrow\rangle \langle \downarrow \!\!|$.
However, for $(\omega,\Delta,\gamma_0)=(1/3,0.2/3,1)$, the final state is quite close to the down-spin state, 
which cannot be captured within the SU(2) FP-TWA. 
In particular, the deviations from the exact result of $\langle {\hat \sigma}_{z}(t) \rangle$ are significant in a time regime 
for $t \gtrsim 2$ due to the approximation error in the SU(2) FP-TWA.
This can be compared to the boson case treated in Sec.~\ref{exact}, suggesting that the failure to describe the vacuum state 
in Fig.~\ref{fig:singlesite}(b) is not specific to the Bose-Hubbard system. 
We note that the {\it density matrix} representation of the Langevin equation adopted here can be equivalently translated into 
a classical Schwinger boson or {\it wave-function} representation with complex number fields, as used in Ref.~\cite{huber2021phase}.

\section{
Base matrices of the SU(3) group
}
\label{def:matrices}

As a complete set of SU(3) base matrices $\{T_{\alpha}\}_{\alpha=1}^8$, in this study, we employ the conventional notation 
for the explicit matrix form, which was also adopted in Refs.~\cite{davidson2015s,nagao20213}, i.e., 
\begin{align}
T_1 
&= \frac{1}{\sqrt{2}}
\begin{bmatrix}
0 & 1 & 0 \\
1 & 0 & 1 \\
0 & 1 & 0
\end{bmatrix},
\;\;\;
T_2 = \frac{1}{\sqrt{2}}
\begin{bmatrix}
0 & -i & 0 \\
i & 0 & -i \\
0 & i & 0
\end{bmatrix}, \nonumber \\
T_3
&=
\begin{bmatrix}
1 & 0 & 0 \\
0 & 0 & 0 \\
0 & 0 & -1
\end{bmatrix},
\;\;\;\;\;\;
T_4
=
\begin{bmatrix}
0 & 0 & 1 \\
0 & 0 & 0 \\
1 & 0 & 0
\end{bmatrix},  \\
T_5
&=
\begin{bmatrix}
0 & 0 & -i \\
0 & 0 & 0 \\
i & 0 & 0
\end{bmatrix},
\;\;\;\;\;\;\;
T_6
= \frac{1}{\sqrt{2}}
\begin{bmatrix}
0 & -1 & 0 \\
-1 & 0 & 1 \\
0 & 1 & 0
\end{bmatrix}, \nonumber \\
T_7 
&= \frac{1}{\sqrt{2}}
\begin{bmatrix}
0 & i & 0 \\
-i & 0 & -i \\
0 & i & 0
\end{bmatrix},
\;\;\;
T_8
= \frac{1}{\sqrt{3}}
\begin{bmatrix}
-1 & 0 & 0 \\
0 & 2 & 0 \\
0 & 0 & -1
\end{bmatrix}. \nonumber
\end{align}
Here, we assume that the highest and lowest weights of $T_3$ correspond to the local vacuum state $|0\rangle$ 
and the doubly occupied state $|2\rangle$, respectively.

\section{
Higher-order corrections and differential equations in the dTWA
}
\label{app:dtwa}

In this appendix, we provide details on the derivation of the dTWA for SU(3) spin systems.
Related discussions specifically for SU(2) spins can be found in Refs.~\cite{schachenmayer2015many,kunimi2021performance}.

Let us introduce one-site reduced operators ${\mathscr A}_{j}^{(1)}(t) = {\rm Tr}_{\underline j} {\mathscr A}^{\otimes N}_t$ 
for the phase-point operator acting on the whole system.
Here, $\underline j$ denotes a trace operation over all sites except for site $j$.
In parallel to this, two-site reduced operators, three-site reduced operators, and higher-rank reduced operators are successively 
defined as ${\mathscr A}_{j,j'}^{(2)} (t)= {\rm Tr}_{{\underline j},{\underline j}'} {\mathscr A}^{\otimes N}_t$, 
${\mathscr A}_{j,j',j''}^{(3)} (t)= {\rm Tr}_{{\underline j},{\underline j}',{\underline j}''} {\mathscr A}^{\otimes N}_t$, and so on.
The time-evolution equation of ${\mathscr A}_{j}^{(1)}(t)$ is derived by tracing the $N-1$ site indices in Eq.~(\ref{eq: eom_ppo}).
However, it is not a closed equation for ${\mathscr A}_{j}^{(1)}(t)$ itself.
Indeed, some inseparable correlations in ${\mathscr A}_{j,j'}^{(2)}(t)$, 
i.e., ${\mathscr B}_{j,j'}^{(2)} (t) \equiv {\mathscr A}_{j,j'}^{(2)}(t) - {\mathscr A}_{j}^{(1)}(t) {\mathscr A}_{j'}^{(1)}(t)$, 
may appear in the determination.
Similarly, to update the values of ${\mathscr A}_{j,j'}^{(2)}(t)$, the information of ${\mathscr A}_{j,j',j''}^{(3)}(t)$ is necessary, 
and more generally, updating ${\mathscr A}_{j,j',\cdots}^{(s-1)}(t)$ requires ${\mathscr A}_{j,j',j'',\cdots}^{(s)}(t)$ at the past step.
This hierarchy is known as the Bogoliubov--Born--Green--Kirkwood--Yvon (BBGKY) hierarchy in the dynamics of multipoint 
correlation functions~\cite{pucci2016simulation}.

The dTWA is formulated as a first-order truncation method of this hierarchy within the single-site decoupling approximation.
In this approximation, each two-site operator is approximated to a factorized operator 
${\mathscr A}^{(2)}_{j, j'}(t)={\mathscr A}^{(1)}_{j}(t){\mathscr A}^{(1)}_{j'}(t)$, i.e., ${\mathscr B}_{j,j'}^{(2)}(t) = 0$.
Any higher-order operator also factorizes.
Hence, the time evolution of the one-site operators can be determined in a closed manner.
This hierarchy truncation is equivalent to the ansatz assumed in Sec.~\ref{subsec: FPTWA}.
A single-site update of the state is performed by integrating the classical equation of motion for single-site variables 
for a finite interval of time.
Beyond this leading-order treatment, it is also possible to incorporate inseparable multisite reduced operators 
step by step in an explicit manner.
However, the computational cost to handle the corresponding time-evolution equations increases exponentially 
as the number of indices involved in an inseparable operator increases.

Notice that, reflecting the trace-conserving property of the time-evolution generator in the master equation, which implies that 
$\langle {\hat 1} \rangle (t) = 1$ holds for all time, the coefficient in front of $I$ remains constant in 
${\mathscr A}_{j,{\rm dTWA}}^{(1)}(t)$.
However, this is not necessarily the case if the nonunitary time evolution belongs to a specific class, 
such as nonunitary dynamics for a non-Hermitian Hamiltonian, which violates the normalization of the wave functions.
A substantial discussion on this issue can be found in Ref.~\cite{vijay2022driven}, where the authors attempt to solve 
the dTWA equation for the 
single-site operator by adopting the quantum jump method, which requires semiclassical evaluations of quantum trajectories 
produced by a non-Hermitian Hamiltonian.

For the case of no dissipation, single-site updates are unitary and governed by the Hamilton equation involving a total of 
$8 L^d$ variables, where $d = 3$.
The explicit form is given as 
\begin{widetext}
\begin{align}
\hbar{\dot r}^{(j)}_{{\rm cl},1} 
&= \frac{\partial H^{W}_{\rm BH}}{\partial r^{(j)}_{{\rm cl},2}}r^{(j)}_{{\rm cl},3} + \frac{\partial H^{W}_{\rm BH}}{\partial r^{(j)}_{{\rm cl},4}}r^{(j)}_{{\rm cl},7} + \frac{\partial H^{W}_{\rm BH}}{\partial r^{(j)}_{{\rm cl},6}}r^{(j)}_{{\rm cl},5} + \sqrt{3}\frac{\partial H^{W}_{\rm BH}}{\partial r^{(j)}_{{\rm cl},7}}r^{(j)}_{{\rm cl},8} - (\text{cyclic perm.}), \nonumber \\
\hbar{\dot r}^{(j)}_{{\rm cl},2} 
&= \frac{\partial H^{W}_{\rm BH}}{\partial r^{(j)}_{{\rm cl},3}}r^{(j)}_{{\rm cl},1} +  \frac{\partial H^{W}_{\rm BH}}{\partial r^{(j)}_{{\rm cl},4}}r^{(j)}_{{\rm cl},6} + \frac{\partial H^{W}_{\rm BH}}{\partial r^{(j)}_{{\rm cl},5}}r^{(j)}_{{\rm cl},7} + \sqrt{3} \frac{\partial H^{W}_{\rm BH}}{\partial r^{(j)}_{{\rm cl},8}}r^{(j)}_{{\rm cl},6} - (\text{cyclic perm.}), \nonumber \\
\hbar{\dot r}^{(j)}_{{\rm cl},3} 
&= \frac{\partial H^{W}_{\rm BH}}{\partial r^{(j)}_{{\rm cl},1}}r^{(j)}_{{\rm cl},2} + \frac{\partial H^{W}_{\rm BH}}{\partial r^{(j)}_{{\rm cl},6}}r^{(j)}_{{\rm cl},7} + 2 \frac{\partial H^{W}_{\rm BH}}{\partial r^{(j)}_{{\rm cl},4}}r^{(j)}_{{\rm cl},5} - (\text{cyclic perm.}), \nonumber \\
\hbar{\dot r}^{(j)}_{{\rm cl},4} 
&= \frac{\partial H^{W}_{\rm BH}}{\partial r^{(j)}_{{\rm cl},7}}r^{(j)}_{{\rm cl},1} + \frac{\partial H^{W}_{\rm BH}}{\partial r^{(j)}_{{\rm cl},6}}r^{(j)}_{{\rm cl},2} + 2 \frac{\partial H^{W}_{\rm BH}}{\partial r^{(j)}_{{\rm cl},5}}r^{(j)}_{{\rm cl},3} - (\text{cyclic perm.}), \nonumber \\
\hbar{\dot r}^{(j)}_{{\rm cl},5} 
&= \frac{\partial H^{W}_{\rm BH}}{\partial r^{(j)}_{{\rm cl},1}}r^{(j)}_{{\rm cl},6} + \frac{\partial H^{W}_{\rm BH}}{\partial r^{(j)}_{{\rm cl},7}}r^{(j)}_{{\rm cl},2} + 2 \frac{\partial H^{W}_{\rm BH}}{\partial r^{(j)}_{{\rm cl},3}}r^{(j)}_{{\rm cl},4} - (\text{cyclic perm.}), \nonumber \\
\hbar{\dot r}^{(j)}_{{\rm cl},6} 
&= \frac{\partial H^{W}_{\rm BH}}{\partial r^{(j)}_{{\rm cl},5}}r^{(j)}_{{\rm cl},1} + \frac{\partial H^{W}_{\rm BH}}{\partial r^{(j)}_{{\rm cl},2}}r^{(j)}_{{\rm cl},4} + \frac{\partial H^{W}_{\rm BH}}{\partial r^{(j)}_{{\rm cl},7}}r^{(j)}_{{\rm cl},3} + \sqrt{3}\frac{\partial H^{W}_{\rm BH}}{\partial r^{(j)}_{{\rm cl},2}}r^{(j)}_{{\rm cl},8} - (\text{cyclic perm.}), \nonumber \\
\hbar{\dot r}^{(j)}_{{\rm cl},7} 
&= \frac{\partial H^{W}_{\rm BH}}{\partial r^{(j)}_{{\rm cl},1}}r^{(j)}_{{\rm cl},4} + \frac{\partial H^{W}_{\rm BH}}{\partial r^{(j)}_{{\rm cl},2}}r^{(j)}_{{\rm cl},5} + \frac{\partial H^{W}_{\rm BH}}{\partial r^{(j)}_{{\rm cl},3}}r^{(j)}_{{\rm cl},6} + \sqrt{3}\frac{\partial H^{W}_{\rm BH}}{\partial r^{(j)}_{{\rm cl},8}}r^{(j)}_{{\rm cl},1} - (\text{cyclic perm.}), \nonumber \\
\hbar{\dot r}^{(j)}_{{\rm cl},8} 
&= \sqrt{3} \frac{\partial H^{W}_{\rm BH}}{\partial r^{(j)}_{{\rm cl},1}}r^{(j)}_{{\rm cl},7} + \sqrt{3} \frac{\partial H^{W}_{\rm BH}}{\partial r^{(j)}_{{\rm cl},6}}r^{(j)}_{{\rm cl},2} - (\text{cyclic perm.}). \nonumber 
\end{align}
\end{widetext}
Here, the derivatives of the Hamiltonian read as
\begin{align}
\frac{\partial H^{W}_{\rm BH}}{\partial r^{(j)}_{{\rm cl},1}} 
&=  - p_1 J \sum_{{\bm \mu}}^{\text{cubic}} ( \langle {\tilde a}^{\dagger}_{j + {\bm \mu}} \rangle +  \langle {\tilde a}_{j + {\bm \mu}} \rangle ),  \nonumber \\
\frac{\partial H^{W}_{\rm BH}}{\partial r^{(j)}_{{\rm cl},2}} 
&= - p_1 J \sum_{{\bm \mu}}^{\text{cubic}} i (\langle {\tilde a}^{\dagger}_{j + {\bm \mu}} \rangle - \langle {\tilde a}_{j + {\bm \mu}} \rangle), \nonumber \\
\frac{\partial H^{W}_{\rm BH}}{\partial r^{(j)}_{{\rm cl},3}}  
&= -\frac{U}{2}, \nonumber \\
\frac{\partial H^{W}_{\rm BH}}{\partial r^{(j)}_{{\rm cl},6}} 
&= - p_2 J \sum_{{\bm \mu}}^{\text{cubic}} ( \langle {\tilde a}^{\dagger}_{j + {\bm \mu}} \rangle +  \langle {\tilde a}_{j + {\bm \mu}} \rangle ), \nonumber \\
\frac{\partial H^{W}_{\rm BH}}{\partial r^{(j)}_{{\rm cl},7}} 
&= - p_2 J \sum_{{\bm \mu}}^{\text{cubic}} i (\langle {\tilde a}^{\dagger}_{j + {\bm \mu}} \rangle - \langle {\tilde a}_{j + {\bm \mu}} \rangle), \nonumber  \\
\frac{\partial H^{W}_{\rm BH}}{\partial r^{(j)}_{{\rm cl},8}} 
&= -\frac{U}{2\sqrt{3}}, \;\;\; \frac{\partial H^{W}_{\rm BH}}{\partial r^{(j)}_{{\rm cl},4}} = \frac{\partial H^{W}_{\rm BH}}{\partial r^{(j)}_{{\rm cl},5}} = 0. \nonumber
\end{align}
The sum with respect to ${\bm \mu}$ is performed over the unit vectors of the cubic lattice 
with ${\bm \mu} \in \{(\pm 1,0,0), (0,\pm 1,0), (0,0,\pm 1)\}$.
We have introduced a simplified notation, e.g., $\langle {\tilde a}_{j + {\bm \mu}}  \rangle = p_1 r^{(j+{\bm \mu})}_{1} + i p_1 r^{(j+{\bm \mu})}_{2} + p_2 r^{(j+{\bm \mu})}_{6} + i p_2 r^{(j+{\bm \mu})}_{7}$, which represents a mean field on the site $j + {\bm \mu}$, and 
can be locally evaluated. Each site $j$ interacts with nearest-neighbor mean fields in this way. 
Moreover, we have substituted nonzero values of the group structure factor $f_{\alpha\beta\gamma}$, 
calculated as~\cite{nagao20213}
\begin{align}
f_{123} &= f_{147} = f_{165} = f_{246} = f_{257} = f_{367} = 1, \nonumber \\
f_{178} &= f_{286} = \sqrt{3},\;\;\;\;\; f_{345} = 2. \nonumber
\end{align}
These values directly follow from the definition of the base matrices in Appendix~\ref{def:matrices}.

\section{
Numerical sampling scheme for phase-space trajectories
}
\label{app:sample}

As described in Sec.~\ref{subsec: FPTWA}, 
to adequately represent quantum fluctuations embedded in the unit filling Mott-insulating state $|\psi_{\rm Mott}\rangle = \prod_{j}{\hat a}^{\dagger}_{j}|0\rangle = \prod_{j}|1\rangle_{j}$, we use the GDW sampling scheme for SU(3) matrices~\cite{zhu2019generalized,nagao20213}. 
This sampling scheme is based on a projective decomposition of each SU(3) base matrix, i.e.,  
\begin{align}
T_{\alpha} = \sum_{a=1}^{3} \lambda^{(a)}_{\alpha} |\phi_{\alpha}^{(a)}\rangle\langle\phi_{\alpha}^{(a)}|,
\end{align}
where $\lambda^{(a)}_{\alpha}$ is the $a$th eigenvalue of $T_{\alpha}$ and $|\phi_{\alpha}^{(a)}\rangle$ 
is the corresponding eigenstate. 
The probability to find ${\hat X}^{(j)}_{\alpha}$ in $\lambda^{(a)}_{\alpha}$ is given by 
$p_{\alpha,a}^{(j)} = | \langle \phi^{(a)}_{\alpha} | 1 \rangle_{j} |^2 \geq 0$.
Since $T_3$ and $T_8$ are already diagonal in our definition (see Appendix~\ref{def:matrices}), 
the classical variables $r_{3}^{(j)}$ and $r_{8}^{(j)}$ take 
$0$ and $2/\sqrt{3}$, respectively, with a probability of $1$. 
In addition,  $r_{4}^{(j)}$ and $r_{5}^{(j)}$, which correspond to nondiagonal matrices, are zero also with a probability of $1$.
However, the other variables, i.e., $r_{1}^{(j)}$, $r_{2}^{(j)}$, $r_{6}^{(j)}$, and $r_{7}^{(j)}$, discretely fluctuate over binary values 
in $\{-1,+1\}$ with an equal probability of $1/2$.
Therefore, the local joint probability distribution for the variables, which provides the proper discrete Wigner function for the local Fock 
vector, reads ${\mathscr W}^{[j]}_{1}(r_1^{(j)},\cdots,r_8^{(j)}) = \delta(r_3^{(j)}) \delta(r_4^{(j)}) \delta(r_5^{(j)}) \delta(r_8^{(j)} - 2/\sqrt{3})\prod_{s=1,2,6,7}[\frac{1}{2}\delta(r_s^{(j)}-1)+\frac{1}{2}\delta(r_s^{(j)}+1)]$.
We can easily check that this probability distribution gives the correct coefficient in the expansion of $|1\rangle \langle 1|$ with respect 
to ${\mathscr A}^{[j]}(r_{1}^{(j)},\cdots,r_{8}^{(j)})$.
A physical understanding of this probability distribution is that the fluctuations of the four variables describe the phase fluctuations 
in the Mott-insulating phase.

The same discrete Wigner function can also be utilized to perform the numerical sampling in the continuous FP-TWA, 
as described in Appendix~\ref{app:fptwa}.
Although Gaussian Wigner functions for the exact continuous Wigner function are typically used for approximate sampling in spin 
systems~\cite{davidson2015s,wurtz2018cluster,nagao20213}, 
the GDW scheme also provides an alternative approach without approximating spin correlations. 
This method has some advantages, such as simplicity of implementation  in numerics and applicability to an arbitrary direct product 
state within positive-definite probabilities.
Moreover, it has been demonstrated in Ref.~\cite{nagao20213} that the results consistent with those using the Gaussian Wigner 
functions are obtained in the SU(3) TWA based on the GDW scheme. 
Therefore, in our study, the GDW scheme is chosen as an efficient initializer of Wiener trajectories in the FP-TWA, 
rather than the Gaussian Wigner scheme.

\section{
Details of the Fokker-Planck TWA for the SU(3) variables
}
\label{app:fptwa}

In the approach adopted to derive the FP equation in Sec.~\ref{subsec: FPTWA}, each SU(3) spin operator is replaced 
with a sum of a classical number and a first-order differential operator~\cite{wurtz2018cluster}, i.e.,
\begin{align}
{\hat X}_{\alpha}^{(j)} \rightarrow x_{\alpha}^{(j)} + \frac{i}{2}f_{\alpha\beta\gamma}x_{\gamma}^{(j)}\frac{\overrightarrow \partial}{\partial x_{\beta}^{(j)}}. \label{eq: approx. bopp}
\end{align}
Here, the differential operator, which can be referred to as a Bopp operator, acts on the Wigner function 
$W_{\rm eff} \equiv ({\hat \rho}_{\rm eff})_{W}$, defined as the continuous Weyl symbol of ${\hat \rho}_{\rm eff}$.
The first term in Eq.~(\ref{eq: approx. bopp}) gives the classical limit of the operator ${\hat X}^{(j)}_{\alpha}$, 
i.e., the Weyl symbol of the SU(3) spin operator given as $x_{\alpha}^{(j)} = ({\hat X}_{\alpha}^{(j)})_{W}$.
The second term in Eq.~(\ref{eq: approx. bopp}), in the form of a differential operator, emerges beyond the classical limit 
to restore the noncommutativity between the spin operators.
Indeed, Eq.~(\ref{eq: approx. bopp}) implies that the commutator between ${\hat X}_{\alpha}^{(j)}$ and ${\hat X}_{\beta}^{(j)}$ 
is transformed as $[{\hat X}_{\alpha}^{(j)},{\hat X}_{\beta}^{(j)}] \rightarrow if_{\alpha \beta \gamma}x_{\gamma}^{(j)} = \{x_{\alpha}^{(j)},x_{\beta}^{(j)}\}_{\rm P.B.}$, 
and hence the amount of noncommutativity is quantified in the phase space by the corresponding Poisson bracket 
for the classical variables.
The Bopp operator given above ignores higher-order derivatives beyond the leading-order contribution of the first-order derivatives, 
thus allowing one to readily derive the SU(3) TWA for open systems as well as isolated systems.

Using the Bopp operators, we can analytically determine $f_{{\rm PA},\alpha}^{(j)} $ and $g_{\alpha,q}^{(j)}$ in the Langevin equation 
in Eq.~(\ref{eq: langevin}) such that 
\begin{align}
f_{{\rm PA},1}^{(j)} 
&= -\frac{\Gamma_{\rm PA}}{4}\left( x_1^{(j)}  - x_4^{(j)}  x_6^{(j)}  - x_5^{(j)}  x_7^{(j)}  \right), \nonumber \\
f_{{\rm PA},2}^{(j)}  
&= -\frac{\Gamma_{\rm PA}}{4}\left( x_2^{(j)}  + x_4^{(j)} x_7^{(j)}  - x_5^{(j)}  x_6^{(j)}  \right), \nonumber \\
f_{{\rm PA},3}^{(j)}  
&= -\frac{\Gamma_{\rm PA}}{2}\left( 2 x_3^{(j)}  - x_4^{(j)} x_4^{(j)}  - x_5^{(j)} x_5^{(j)} \right), \nonumber \\
f_{{\rm PA},4}^{(j)}  
&= - \frac{\Gamma_{\rm PA}}{2} \left( x_3^{(j)} + 1 \right)x_4^{(j)}, \label{eq:f_PA} \nonumber \\
f_{{\rm PA},5}^{(j)}  
&= - \frac{\Gamma_{\rm PA}}{2} \left( x_3^{(j)}  + 1 \right)x_5^{(j)}, \\
f_{{\rm PA},6}^{(j)}  
&= -\frac{\Gamma_{\rm PA}}{4}\left( x_6^{(j)}  + x_1^{(j)} x_4^{(j)} + x_2^{(j)} x_5^{(j)} \right), \nonumber \\
f_{{\rm PA},7}^{(j)}  
&= -\frac{\Gamma_{\rm PA}}{4}\left( x_7^{(j)}  - x_2^{(j)}  x_4^{(j)}  + x_1^{(j)}  x_5^{(j)}  \right), \nonumber \\
f_{{\rm PA},8}^{(j)}  
&= 0, \nonumber
\end{align}
and 
\begin{align}
g_{1,q=1}^{(j)} 
&= g_{2,q=2}^{(j)} = \frac{\sqrt{\Gamma_{\rm PA}}}{2}x_{6}^{(j)}, \nonumber \\
g_{1,q=2}^{(j)} 
&= - g_{2,q=1}^{(j)} = \frac{\sqrt{\Gamma_{\rm PA}}}{2}x_{7}^{(j)}, \nonumber \\
 g_{3,q=1}^{(j)} 
&= \sqrt{\Gamma_{\rm PA}} x_{4}^{(j)}, \nonumber \\
g_{3,q=2}^{(j)} 
&= \sqrt{\Gamma_{\rm PA}} x_{5}^{(j)}, \nonumber \\ 
g_{4,q=1}^{(j)} 
&= g_{5,q=2}^{(j)} = - \sqrt{\Gamma_{\rm PA}} x_{3}^{(j)}, \label{eq:g} \\
g_{6,q=1}^{(j)} 
&= g_{7,q=2}^{(j)} = -\frac{\sqrt{\Gamma_{\rm PA}}}{2} x_{1}^{(j)}, \nonumber \\
g_{7,q=1}^{(j)} 
&= - g_{6,q=2}^{(j)} = \frac{\sqrt{\Gamma_{\rm PA}}}{2} x_{2}^{(j)}, \nonumber \\
g_{4,q=2}^{(j)} 
&= g_{5,q=1}^{(j)} = g_{8,q=1}^{(j)} = g_{8,q=2}^{(j)} = 0. \nonumber
\end{align}
Notice that $f_{{\rm PA},\alpha}^{(j)}$ are nonlinear functions of the phase-space variables, implying the limitation of the leading order 
truncation in the Bopp operators.

We now explain how the numerical sampling is carried out to compute an average over stochastic Langevin trajectories.
First, we draw a set of random spins from a Wigner distribution.
In this study, we use a generalized discrete Wigner sampling scheme, described in Appendix~\ref{app:sample}, 
for the purpose of sampling.
The initialized spins are then propagated with the Langevin equation in Eq.~(\ref{eq: langevin}) until the final time at $t = t_1 > 0$, 
so that we have a non-differentiable trajectory $ x^{(j)}_{{\rm cl},\alpha}(t_1) = x^{(j)}_{{\rm cl},\alpha}[ t_1; x^{(j)}_{{\rm cl},\alpha}(n \Delta t),\cdots, x^{(j)}_{{\rm cl},\alpha}(\Delta t), x^{(j)}_{{\rm cl},\alpha}(0)]$.
To numerically integrate the Langevin equation, we employ the first-order Euler-Maruyama 
algorithm~\cite{gardiner2009stochastic} with $\Delta t = t_1/(n+1)$, where $n$ is an integer providing a sufficiently small $\Delta t$, 
ensuring the stability of the simulation. 
This trajectory, also referred to as a path of the Wiener process, is a function of generated random numbers 
$dw_{q}^{(j)}(t)$ during the time sequence and is thus not differentiable everywhere.
Using the registered data of this trajectory, we evaluate ${\cal O}_{W} \left[ \{ x^{(j)}_{{\rm cl},\alpha}(t) \} \right]$ 
for each $t \in [0,t_1]$, which is a fluctuating classical function.
We iteratively generate many stochastic trajectories and take the Monte Carlo average of ${\cal O}_{W} \left[ \{ x^{(j)}_{{\rm cl},\alpha}(t) \} \right]$ over these trajectories until the convergence is finally reached.
In contrast to the dTWA case, the FP-TWA must successively generate Gaussian noises at each time step after taking initial noises 
at $t=0$ from the Wigner distribution. 
We should also note that higher-order algorithms for solving the Langevin equation, such as the Platen scheme to second order of 
$\Delta t$~\cite{breuer2002theory}, are quite hard to implement in practice due to the nondifferentiable manner of trajectories.

\bibliography{ref}

\end{document}